\documentclass[prd,twocolumn,showpacs,tightenlines,amssymb,
   eqsecnum,floatfix,superscriptaddress,preprintnumbers]{revtex4}
\usepackage{epsf}
\begin{document}

\newcommand{\scri}{{\cal I}}
\newcommand{\hor}{{\cal H}}
\newcommand{\JM}{{\cal J}^-}
\newcommand{\JP}{{\cal J}^+}
\newcommand{\Norm}[2]{\|#1\|_#2}

\preprint{NSF-KITP-03-49}
\preprint{AEI-2003-053}
\preprint{gr-qc/0306098}
\title{Mode coupling in the nonlinear response of black holes}

\author{Yosef Zlochower}
\affiliation{Department of Physics and Astronomy, University of
Pittsburgh, Pittsburgh Pennsylvania 15260.}

\author{Roberto G\'omez}
\affiliation{Pittsburgh Supercomputing Center, 4400 Fifth Avenue,
Pittsburgh, Pennsylvania 15213.}

\affiliation{Department of Physics and Astronomy, University of
Pittsburgh, Pittsburgh Pennsylvania 15260.}

\author{Sascha Husa}
\affiliation{Albert Einstein Institute, Max Planck Gesellschaft,
Haus 1, Am M\"uhlenberg, Golm, Germany.}

\author{Luis Lehner}
\affiliation{Department of Physics and Astronomy, Louisiana State
University, Baton Rouge, LA 70803}

\author{Jeffrey Winicour}
\affiliation{Albert Einstein Institute, Max Planck Gesellschaft,
Haus 1, Am M\"uhlenberg, Golm, Germany.}

\affiliation{Department of Physics and Astronomy, University of
Pittsburgh, Pittsburgh Pennsylvania 15260.}

\date{\today}

\begin{abstract}

We study the properties of the outgoing gravitational wave produced when
a non-spinning black hole is excited by an ingoing gravitational wave.
Simulations using a numerical code for solving Einstein's equations
allow the study to be extended from the linearized approximation, where
the system is treated as a perturbed Schwarzschild black hole, to the
fully nonlinear regime. Several nonlinear features are found which bear
importance to the data analysis of gravitational waves. When compared to
the results obtained in the linearized approximation, we observe large
phase shifts, a stronger than linear generation of gravitational wave
output and considerable generation of radiation in polarization states
which are not found in the linearized approximation. In terms of a
spherical harmonic decomposition, the nonlinear properties of the
harmonic amplitudes have simple scaling properties which offer an
economical way to catalog the details of the waves produced in such
black hole  processes.

\end{abstract}


\maketitle

\section{Introduction}

A sensitive array of detectors is searching for the gravitational
waves produced in violent astrophysical scenarios.  Detecting and
understanding the information carried by these waves will be crucial
to elucidate many poorly understood phenomena in astrophysics and
cosmology.  In particular, the
formation of black holes and their interaction with the surrounding
media might produce gravitational waves detectable by space or earth
based observatories, such as LISA and LIGO (see for instance
\cite{CutlerThorne,Grishchuck:2003uh,Hughes:2002yy}). It is therefore
important to consider these kinds of scenarios and produce reliable
waveform estimates which can be used as input for data analysis
techniques. 

In the case of LISA, its frequency band will be sensitive to systems
involving supermassive black holes. While the formation process of
these is not well understood, different working models suggest the
following mechanisms: formation by a series of smaller black hole
mergers; growth from smaller holes by accretion or collapse of massive
gas accumulations~\cite{Rees:1999sk}. Irrespective of the mechanism,
after some stage it would lead to the scenario of a massive black hole
which is being strongly perturbed.  For earth based detectors, systems
of interest include mergers of stellar mass black holes and/or neutron
stars. The early stages of these systems can be fairly accurately
described by post-Newtonian approximations (e.g.
\cite{BlanchetPN,DamourPN}), while the late stage can be described by
linearized perturbations off a single black hole spacetime. The
intervening strong field stage can only be handled accurately via
numerical simulations of Einstein equations which, to date, are not
yet available (see \cite{Lehnerreview,Baumgartereview} for a review on
the status of efforts in this direction).

Irrespective of whether the massive black hole formed from accretion of
matter or from mergers of smaller black holes, quite useful information
can be obtained by exploiting the fact that at some stage the system can
be treated as a perturbed, single black hole spacetime (as was
demonstrated in~\cite{pullin1,pullin2}).

Therefore, a numerical code that can stably deal with generic single
black hole spacetimes can be useful in the study of nonlinear
disturbances, probing whether there are robust features which are
reflected in the waveforms produced. If these features are present, the
information can be incorporated in data analysis (see for instance
~\cite{Flanagan:1998sx,Brady:1998ji}), which would be of considerable
value until more refined waveforms can be obtained.

A mature characteristic evolution code, the PITT null code, is capable
of handling generic single black hole spacetimes~\cite{high}. The code
computes the Bondi news function describing radiation at future null
infinity $\scri^+$, which is represented as a finite boundary on a
compactified numerical grid. The Bondi news function is an invariantly
defined complex radiation amplitude $N= N_{\oplus}+i N_{\otimes}$, whose
real and imaginary parts correspond to the time derivatives $\partial_t
h_{\oplus}$ and $\partial_t h_{\otimes}$ of the ``plus'' and ``cross''
polarization modes of the strain incident on a gravitational wave
antenna. An alternative approach to calculating radiation in black hole
spacetimes might be based upon the Cauchy formulation, which has
recently undergone significantly
improvement~\cite{yo,alcubierre,scheel,shoem}. In the present work, we
employ the characteristic code, developed in~\cite{high} and further
refined in~\cite{gomez2001}, to study the nonlinear response to the
scattering of gravitational radiation off a non-spinning black hole (the
case for spinning black holes being deferred for future work). Although
the gravitational waves will be weak when they reach a detector, they
are produced in a region where nonlinear effects are important.

In this work, we investigate in a simple setting how such
nonlinearities affect the produced waveforms. Considerably more work
certainly lies ahead but already important effects are noticed.
Namely, by perturbing the black hole with a pulse with a single
$(\ell,m)$ mode of amplitude $A$ we observe:

\begin{itemize}
\item{There is generation of additional modes whose amplitudes scale
  as precise powers of $A$. This suggests an economical way of producing
  a waveform catalog based upon combining the linearized results with a
  modest number of (non-linear) simulations.}
\item{There is significant phase shifting and frequency modulation in
  the nonlinear waveforms.} 
\item{There is nonlinear amplification of the gravitational wave
  output.}
\item{There is a significant generation of radiation in polarization
  states not present in the linearized approximation.}
\end{itemize}

Throughout this work to illustrate the above mentioned effects  we
standardize the input pulse to ($\ell=2$, $m=0$) and ($\ell=2$, $m=2$)
quadrupole modes, although the simulations can be carried out with any
input data. Our main objective here is to demonstrate the
effectiveness of the characteristic code in providing a complete
analysis of the mode coupling in the outgoing waveform.  The
robustness observed in this simple case opens the door to performing
in-depth analysis of more general input pulses. For example, a
localized ingoing pulse of gravitational energy, patterned after the
distortion in the gravitational  field around a localized distribution
of matter, could be used to mimic the  effects of the infall of matter
onto a black hole until more realistic  calculations are achievable.

Mode mixing and the transition from the linear to the nonlinear regime
have previously been studied by Allen et al.~\cite{Allen} and Baker et
al.~\cite{Baker:2000np} in a Cauchy evolution framework. They extract
nonlinear waveforms by matching the perturbative and nonlinear
solutions on a worldtube at $r=15M$, where $M$ is the unperturbed
black hole mass. Our work is complementary to theirs in the sense that
we use a characteristic formalism to carry out the evolutions. A
characteristic approach offers greater flexibility and control in
prescribing initial data. In the Cauchy approach, elliptic constraints
must be solved in order to provide initial data. In order to simplify
the constraint problem, Allen et al. \cite{Allen} use time-symmetric
initial data, which intrinsically contain equal amounts of ingoing and
outgoing radiation. In our characteristic approach, we prescribe
initial data on a pair of intersecting null hypersurfaces, one of
which is ingoing and the other outgoing. Data on each null
hypersurface can be prescribed freely to represent a pulse with
arbitrary waveform. Furthermore, the data on an outgoing null
hypersurface can be identified with an ingoing wave; and data on an
ingoing null hypersurface, with an outgoing wave. There is no
comparable way to prescribe purely ingoing or outgoing Cauchy data. 

A characteristic treatment of mode mixing in the axisymmetric case
($m=0$) has previously been undertaken by Papadopoulos~\cite{papad},
using the initial axisymmetric version of the PITT code. Papadopoulos
carried out an illuminating study of  the propagation of an outgoing
pulse by evolving it along a family of ingoing null hypersurfaces with
outer boundary at $r=60M$.  The evolution is stopped before the pulse
hits the outer boundary in order to avoid spurious reflection effects
and the radiation is inferred from data at $r=20M$. As in the Cauchy
treatment of Ref's~\cite{Allen,Baker:2000np}, gauge ambiguities arise
from reading off the radiation waveform at a finite radius.

\begin{figure}
  \centerline{\epsfxsize=3in\epsfbox{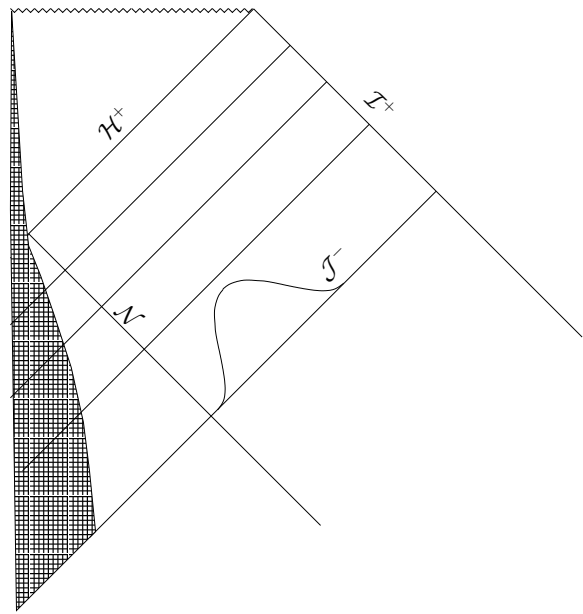}}
  \caption{ The physical setup for the scattering problem. A
            star of mass $M$ has undergone spherically symmetric
            collapse to form a black hole. The ingoing null worldtube
            ${\cal N}$ lies outside the collapsing matter.  The metric
            inside ${\cal N}$ (but outside the matter) is vacuum
            Schwarzschild. Outside of ${\cal N}$, data for an ingoing
            pulse is specified on the initial outgoing null hypersurface
            $\JM$. As the pulse propagates to the black hole event
            horizon $\hor^+$, part of its energy is scattered to
            $\scri^+$.}
\label{fig:setup}
\end{figure}

In the present work, we study the scattering of both axisymmetric and
non-axisymmetric ingoing pulses by evolving along outgoing null
hypersurfaces.  The physical setup is described in
Fig.~\ref{fig:setup}. The outgoing null hypersurfaces extend to future
null infinity $\scri^+$ on a compactified numerical grid.
Consequently, there is no need for either an artificial outer boundary
condition or an interior extraction worldtube. The outgoing radiation
is computed in the coordinates of an observer in an inertial frame at
infinity, thus avoiding any gauge ambiguity in the waveform. Although
the technical setup is very different from the previous
treatments~\cite{Allen,Baker:2000np,papad}, the simulations display
several qualitatively similar features (see Sec.~\ref{sec:concl}).

In Sec.~\ref{sec:spin}, we briefly summarize the characteristic
formalism on which the paper is based. In Sec.~\ref{sec:prob}, we give
the geometrical setup for initializing and setting boundary conditions
for the simulations. In Sec.~\ref{sec:calibration}, we calibrate the
accuracy of the full nonlinear code against results obtained with
independent linearized characteristic codes~\cite{close1,close2} that
solve the Teukolsky equations~\cite{Teukolsky72,Teukolsky73} for
perturbations about Schwarzschild spacetime.  We show, in the regime
where linear perturbation theory is valid, that the full nonlinear code
reproduces the Bondi news function obtained from the perturbative codes
to second order accuracy. 

Simulations of nonlinear mode-mode coupling are described in
Sec.~\ref{sec:coupling}. The nonlinear effects on the waveform of the
outgoing radiation can be cleanly separated out by comparison with the
waveform produced by the linearized code. The simulations reveal several
nonlinear features of the waveform that have potential importance to
the design of templates for data analysis. These are summarized in
Sec.~\ref{sec:concl}.

\section{The nonlinear scattering problem}
\label{sec:spin}

Following~\cite{cce,high,gomez2001,Gomez:2002ev}, we use coordinates
based upon a family of outgoing null hypersurfaces, letting $u$ label
these hypersurfaces, $x^A$ $(A=2,3)$, label the null rays, and $r$ be
a surface area coordinate, such that in the $x^\alpha=(u,r,x^A)$
coordinates the metric takes the Bondi-Sachs form~\cite{bondi,sachs}

\begin{eqnarray}
   ds^2 & = & -\left[e^{2\beta}\left(1 + {W \over r}\right)
              - r^2h_{AB}U^AU^B\right]du^2
              - 2e^{2\beta}dudr
 \nonumber \\
            & - & 2r^2 h_{AB}U^Bdudx^A
              + r^2h_{AB}dx^Adx^B.    \label{eq:bmet}
\end{eqnarray}
Here $W$ is related to the Bondi-Sachs variable $V$ by $V=r+W$ and
$det(h_{AB})=det(q_{AB})$, with $q_{AB}$ a standard unit sphere
metric. We represent $q_{AB}$ in terms of a complex dyad $q_A$
satisfying $q^Aq_A=0$, $q^A\bar q_A=2$, $q^A=q^{AB}q_B$, with
$q^{AB}q_{BC}=\delta^A_C$ and $ q_{AB} =\frac{1}{2}\left(q_A \bar
q_B+\bar q_Aq_B\right)$. The angular coordinates $x^A=(q,p)$ used in
the code are based upon a complex stereographic coordinate $z=q+ip$
for the unit sphere metric $$q_{AB}dx^A dx^B = \frac
{4}{1+q^2+p^2}(dq^2+dp^2),$$ with two patches used to cover the
sphere. In the north patch, the stereographic coordinate is related to
standard spherical coordinates $(\theta,\phi)$ by $z_N = \tan
\frac{\theta}{2} e^{i \phi}$ (see App.~\ref{ap:spin}).

We represent tensors on the sphere by spin-weighted
variables~\cite{competh}.  Thus the conformal metric $h_{AB}$ is
represented by the complex function $J=\frac{1}{2}h_{AB}q^Aq^B$, and
the real function $K=\frac{1}{2}h_{AB}q^A\bar q^B$, where $K^2=1+J\bar
J$ on account of the condition $\det(h_{AB})=\det(q_{AB})$. The metric
functions $U^A$ are similarly encoded in the complex function $U=q_A
U^A$~\cite{cce,high}.  Angular derivatives are expressed in terms of
$\eth$ and $\bar\eth$ operators. A brief description, along with our
conventions for spin-weighted spherical harmonics, is given in
App.~\ref{ap:spin}. For details, see Ref.~\cite{competh}.

The vacuum Einstein equations form (i) a hierarchy of null
hypersurface equations for the radial derivatives of  $\beta$, $U$ and
$W$ (and also for auxiliary variables $\nu$, $k$, $B$ and $Q$ used to
reduce the equations to first order differential form in the angular
coordinates), (ii) a complex evolution equation for the metric
function $J$ and (iii) a set of subsidiary equations which are
satisfied by virtue of the other equations if they are satisfied on a
null or timelike worldtube.  In our case, a null worldtube forms the
inner boundary of the computational domain, and we solve the equations
induced there in the form discussed in  Ref.~\cite{compnull}.  Most of
the discussion in the present paper will deal with an
``ingoing''pulse, in which case Schwarzschild data are posed on the
inner boundary, the subsidiary equations are then solved
automatically.  The explicit form of the equations (i) and (ii) is
given in Eq's.~(16)--(26) of Ref.~\cite{gomez2001}.

Given initial data $J$ on an outgoing null hypersurface extending to
$\scri^+$ and boundary data on an inner timelike or null world tube,
the PITT null code carries out the future evolution of the spacetime.
The evolution extends to $\scri^+$ where the Bondi news function $N$
is calculated. In the present work, the computation of the news is
carried out with enhanced accuracy by the new method described in
App.~\ref{sec:news}.

\section{Data for the scattering problem}
\label{sec:prob}

The nonlinear simulations described in Sec.~\ref{sec:coupling} model
the scattering of an ingoing pulse in the region exterior to a
spherically symmetric collapsing star of mass $M$, as depicted in
Fig.~\ref{fig:setup} (the mass $M$ will always be scaled to 1 in our
simulations). The free initial data for the pulse consist of the
metric function $J$ on the initial outgoing null hypersurface, denoted
by $\JM$. In addition, Schwarzschild null data is given on the
interior null hypersurface ${\cal N}$, which is causally unaffected by
the pulse.  The evolution code then provides the news function at
$\scri^+$.  For computational simplicity, we simulate a mathematically
equivalent problem by considering the Kruskal continuation of the
Schwarzschild interior so that the inner boundary is extended to the
ingoing branch of the $r=2M$ null hypersurface, which we denote by
$\hor^-$.  In our metric ansatz (\ref{eq:bmet}) Schwarzschild data on
$\hor^-$ correspond to setting $\beta = 0$, $U = 0$, $W = -2 M$.  A
schematic view of this setup, in the linear regime, is provided in
Fig.~\ref{fig:coord_sing} (Fig.~\ref{fig:coord_sing} shows the
compactified exterior Kruskal quadrant which includes $\scri^-$ as
well as $\scri^+$). In the nonlinear simulations, the data on $\hor^-$
is set to the Schwarzschild data $J=0$, which implies that the
intended Schwarzschild data is also induced on ${\cal N}$. In the code
calibration tests considered in Sec.~\ref{sec:calibration} we also
consider outgoing pulses generated by data on $\hor^-$.

We used two different types of evolution coordinates, $u$ and $\hat
u$.  For the calibration runs we chose as our evolution coordinate $u$
which is an affine parameter along the null generators of $\hor^-$.
For the remaining runs we chose as our evolution coordinate $\hat u$
which is the standard Schwarzschild retarded time on $\hor^-$ ($\hat u
= t - r^*$).  Later, when analyzing gravitational wave signals, we
will use Bondi time -- an affine parameter $\tilde u$ along the null
generators of $\scri^+$.  For Schwarzschild, Bondi time coincides with
the standard Schwarzschild retarded time $\tilde u=\hat u$, and $u=-M
exp(-\tilde u/4M)$.

We prescribe the input data for the pulse in the form $J = f(x)
\,{}_2R_{\ell\,m}$, where $x = r$ for initial data on $\JM$ or $x = u$
for data on $\hor^-$ and ${}_sR_{\ell\,m}$ are spin-weighted spherical
harmonics with real potentials (see App.~\ref{ap:spin}).  The
${}_sR_{\ell\,m}$ are linear combinations of the standard
${}_sY_{\ell\,m}$ and form a complete basis for smooth spin-weight $s$
functions. In linear theory a particular ${}_2R_{\ell\,m}$ mode in the
data produces a Bondi news function that contains only that mode.
Conversely, data constructed from the standard spin-weighted spherical
harmonics ${}_2Y_{\ell\,m}$ would produce both the ${}_2Y_{\ell\,m}$
and ${}_2Y_{\ell\,-m}$ modes in the Bondi news.  An explanation for
this linear mode-coupling and explicit formulae for the
${}_sR_{\ell\,m}$ are  given in App.~\ref{ap:spin}. 
 
The function $f$ determines the radial profile of the pulse. As for
its initial angular dependence, we restrict our attention here to
input data with $\ell=2$ and $m=0$ or $m=2$. In the regime where the
linear approximation is valid, the Bondi news function $N$ will also
be of this form but nonlinear effects lead to the presence of higher
order modes.  Parity and reflection symmetry implies that only even
$\ell$ and even $m\geq 0$ modes be generated by such nonlinear
effects.

The input  pulse is specified in terms of the function 
\begin{equation}
f(x) =  A\frac{2^{2n}
    (x-x_{min})^n(x_{max}-x)^n}{(x_{max}-x_{min})^{2 n}}
\label{eq:data}
\end{equation}
where $A$ controls the amplitude and $n$ controls the steepness.  For
outgoing pulses, we prescribe boundary data on $\hor^-$ with the
specific profile
\begin{equation}
   J|_{\hor^-}(u,x^A) = f(u)
   \,{}_2R_{2\,m},
\label{eq:dataout}
\end{equation}
with $n=6$, for $u_{min} < u < u_{max}$; elsewhere we set $J=0$.  For
an ingoing pulse, we give data on the initial outgoing null
hypersurface of the form
\begin{equation}
  J(r, x^A) = f(x) 
  \,{}_2R_{2\,m}
  \label{eq:datain}
\end{equation}
for $x_{min} <x< x_{max}$,  where $x = r/(2M +r)$, with either $n=3$
or $n=6$; elsewhere we set $J=0$. We also set $J=0$ on the inner
boundary $\hor^-$.

\section{Calibration of the nonlinear code against perturbative
  solutions}
\label{sec:calibration}

We calibrate the Bondi news function obtained from the PITT nonlinear
code against values obtained from a characteristic perturbative
code~\cite{close1,close2} based upon the Teukolsky equations for the
Newman-Penrose quantities~\cite{np} 
$\tilde \psi_0 = C_{a b c d} \tilde l^a m^b \tilde l^c m^d$ and 
$\tilde \psi_4 = C_{a b c d} \tilde n^a \bar m^b \tilde n^c \bar m^d$.

We fix the null tetrad by setting $\tilde l_a = -\nabla_a \tilde u$
and $m^a = q^a/\sqrt{2} r$, where $\tilde u = t - r^*$ in the
background Schwarzschild spacetime [$r^*=r + 2M \log(r/2M-1)$ is the
Regge-Wheeler tortoise coordinate]. Rather than evolving the
spin-weighted quantities $\tilde \psi_0$ and $\tilde \psi_4$, we
evolve equivalent spin-zero potentials (see App.~\ref{ap:spin})
rescaled by appropriate factors of $r$. These evolution variables are
$\tilde F_4$, where $\bar\eth^2 \tilde F_4 = r \tilde \psi_4$, and
$\hat F_0$, where $\eth^2 \hat F_0 = (1-2 M/r)^2 r^5 \tilde \psi_0$.
Both $\tilde F_4$ and $\hat F_0$ are constructed to be finite on
$\scri^+$.

The perturbative variable $\tilde F_4$ has a complicated dependence on
the Bondi metric variables. However, on $\hor^-$, where the data for
the outgoing pulse is prescribed, the expression for $\tilde F_4$
reduces to
\begin{equation}
  \tilde F_4 =\frac{u^2}{16 M}  \bar j_{,uu},
  \label{eq:p4j}
\end{equation}
where $u= -M \exp(-\tilde u/4 M)$ is an affine parameter along the
null generators of $\hor^-$ and $j$ is the spin-zero potential for
$J$, i.e.\ $J=\eth^2 j$. On an outgoing null hypersurface, where the
data for an ingoing pulse is prescribed, the perturbative variable
$\hat F_0$ has the simple dependence on $j$ given by
\begin{equation}
  \hat F_0 =\frac{1}{2} r(r-2M)^2\partial_r(r^2\partial_r j).
  \label{eq:p0j}
\end{equation}
For input data consisting of an outgoing pulse, where we specify
compact support boundary data for $J$ on $\hor^-$ and set $J=0$ on
$\JM$, we directly obtain equivalent boundary data for $\tilde F_4$
via Eq.~(\ref{eq:p4j}). For input consisting of an ingoing pulse,
where we specify the compact support data on $\JM$, we directly obtain
the equivalent boundary data for $\hat F_0$ via Eq.~(\ref{eq:p0j}).
For the calibration tests presented here we choose input data with
($\ell=2$, $m=0$) spherical harmonic dependence for which $\tilde F_4
= f_4(u,r) R_{2\,0}$ [$R_{\ell\,m}={}_0R_{\ell\,m}$] and $\hat F_0 =
f_0(u,r) R_{2\,0}$. The perturbative code provides solutions for $f_4$
and $f_0$. Details of the evolution code for $f_4$ can be found
in~\cite{close1}; and details for $f_0$, in~\cite{close2}.

In the perturbative approach, the Bondi news function is either
obtained directly from $\tilde F_4$ on $\scri^+$ or indirectly from
$\hat F_0$~\cite{close2}. The Bondi news function is given by
\begin{equation}
  \bar N(\tilde u) = \bar\eth^2 \int_{-\infty}^{\tilde u}
    \tilde F_4|_{\scri^+} (\tilde t) d\tilde t,
  \label{eq:NfromF4}
\end{equation}
where the integral in Eq.~(\ref{eq:NfromF4}) is over Bondi time. In
the simulation of an outgoing pulse, the input data for $\tilde F_4$
on $\hor^-$ is set to zero until the start of the evolution at $\tilde
u = \tilde u_{min}$.  In this case we calculate the Bondi news through
a second-order accurate midpoint-rule integration of $\tilde F_4$.

In the simulation of the scattering of an ingoing pulse, the
computation of the news function is more complicated. For the present
case of an ($\ell=2$, $m=0$)  linearized mode, the news function is
related to $\hat F_0$ by
\begin{equation}
   N + \frac{M}{2} \partial_{\tilde u} N =
       \frac{1}{6}\eth^2\partial_{\tilde u}^3 \hat F_0|_{\scri^+}.
\label{eq:NfromF0}
\end{equation}
Care must be taken in initiating the integration of
Eq.~(\ref{eq:NfromF0}) because the Bondi news function $N$ does not
vanish at the initial evolution time. We need an accurate initial
value for $N$, which we obtain by integrating Eq.~(\ref{eq:p0j}) to
obtain the metric variable $J$. In the linear regime, the Bondi news
function can be calculated from the value of $J$ and its radial
derivative at $\scri^+$~\cite{high}.  In the present case of an
($\ell=2$, $m=0$) linearized mode,
\begin{eqnarray}
 N = -\frac{1}{2} \left(3J+r^2 \partial_r\partial_{\tilde u} J
        \right ) |_{\scri^+}
  \label{eq:NfromJ}
\end{eqnarray}
(a generic formula for the linearized News as a function of
$J|_\scri^+$ is given in Eq.~(\ref{eq:NewsCoef})).  We use
Eq.~(\ref{eq:NfromJ}) to obtain the news function on the initial slice
and also to compute it at later times as a check that it produces the
same values as the integration of (\ref{eq:NfromF0}).

Because the perturbative code is one-dimensional, computations on very
large radial grids can be carried out to provide effectively exact
solutions for calibrating the error in the Bondi news function
calculated using the nonlinear code. Such a check must be carried out
in the range of validity of the linear approximation, and the error in
the perturbative calculation must be sufficiently smaller than the
discretization error of the nonlinear null code.  For this purpose,
simulations with the perturbative code were carried out with 16001
radial grid points for the outgoing pulse.  For the ingoing pulses
(for which the perturbative code is less accurate), we used 2501,
5001, and 10001 radial grid points, and then performed a Richardson
extrapolation to obtain high accuracy.

The initial and boundary data given by Eq's.~(\ref{eq:dataout}) and
(\ref{eq:datain}) are compact pulses with relatively steep profiles,
which require comparatively large grid sizes to attain good
resolution.  The specific form of the profiles lead to $C^3$
differentiability of the associated perturbative quantities $\tilde
F_4$ and $\hat F_0$, a requirement which ensures that the news
function can be obtained from the perturbative calculation to second
order accuracy. In all tests we scale the initial mass of the system
to $M=1$. 

\subsection{Propagation of outgoing pulses}
\label{sec:fullpolytest}

In our first battery of tests, which provide stringent tests of the
ability of the code to carry radiation away from the horizon $\hor^-$
to null infinity $\scri^+$, we prescribe an arbitrary perturbation on
the horizon while setting the perturbation of the initial outgoing
null cone to zero. The horizon data for $J$ are given by
Eq.~(\ref{eq:dataout}) with $u_{min} =-1.99$, $u_{max}=-1.5$, and $A =
2.6\cdot 10^{-5}$. This corresponds to a relatively short pulse of
$\Delta \tilde u \approx 1.13 M$ in the background Schwarzschild
retarded time.  The corresponding data for $\tilde F_4$ are obtained
by applying Eq.~(\ref{eq:p4j}) to Eq.~(\ref{eq:dataout}). We measure
the error in the Bondi news function $N_{code}$ with the $L_\infty$
norm $\| N_{code} - N_{pert}\|_{\infty}$ restricted to the interior of
the equator on the stereographic patches.  Our primary concern is to
check convergence and the evolution was stopped at $u=-1.5$ when the
horizon data vanishes. The perturbative news function was obtained
using $16001$ radial grid points and an initial time-step of
$du=8\cdot10^{-5}$.

The nonlinear characteristic code uses a uniform radial grid of $n_x$
points on the compactified coordinate $x=r/(R+r)$, where $R=2M$ is the
initial horizon radius.  The coordinate range is $1/2<x<1$, with
$x=1/2$ at $\hor^-$ and $x=1$ at $\scri^+$.  We introduce two
additional ghost zones inside the horizon in accord with the start-up
algorithm described in~\cite{Gomez:2002ev}. The angular grid is also
uniform, with $n_q$ grid points spanning the range $(-q_S,q_S)$ in
both the $q$ and $p$ directions. We set $q_S =1.2$ in order to provide
a finite overlap between the two stereographic patches. In the
convergence tests we take $n_q = 12 i+5$, $n_x = 30 i +3$, with $i$
taking integer values from 6 to 11.  We keep the time step $du$
constant during each run, at $du = 5.6\cdot 10^{-3} /i$, below the
smallest value required for the CFL condition to be satisfied and
scaling with $i$ to guarantee convergence.

\begin{figure}
  \centerline{\epsfxsize=3.178in\epsfbox{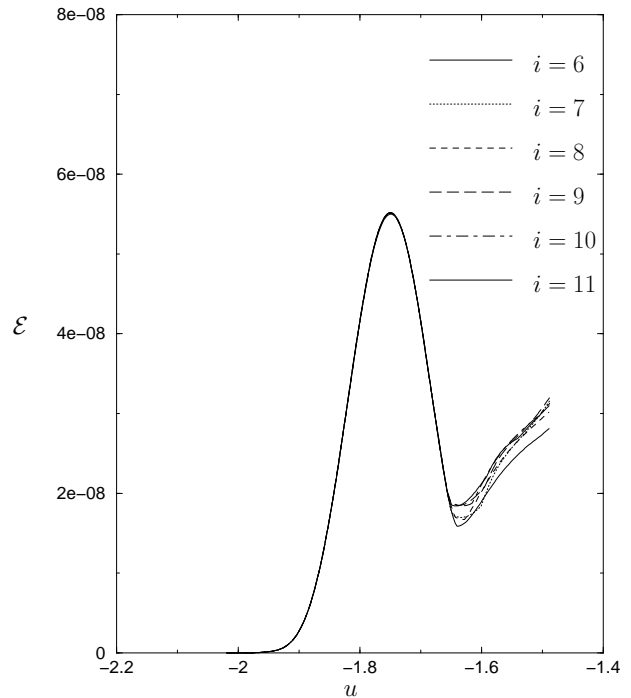}}
  \centerline{\ }
  \caption{ The rescaled $L_\infty$ norms ${\cal E} =
            (i/6)^2 \|N_{code} -N_{pert}\|_{\infty}$ (versus time) of
            the Bondi news for the compact support outgoing pulse.}
  \label{fig:fullpolyerror}
\end{figure}

\begin{figure}
  \centerline{\epsfxsize=3in\epsfbox{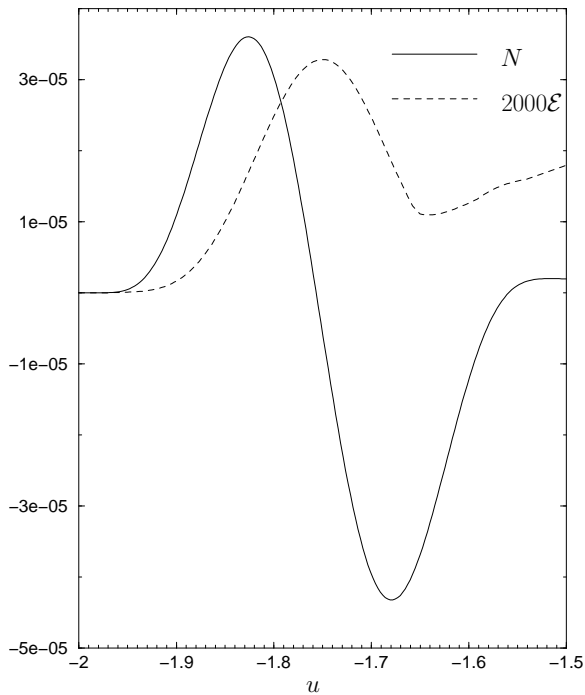}}
  \centerline{\ }
  \caption{ The news and $L_\infty$ norm ${\cal E}=
            \|N_{code} -N_{pert}\|_{\infty}$ (rescaled by a factor of
            2000) for the compact support outgoing pulse.}
  \label{fig:polyanderr}
\end{figure}

Since the code is second order accurate, and the linear approximation
holds, the error norm should be inversely proportional to the square
of the grid size, i.e.\ to $i^2$. Figure~\ref{fig:fullpolyerror} shows
the rescaled $L_{\infty}$ norm of the error for the above 6 grids.
Note the perfect overlap at early times (prior to $u=-1.65$). The
later errors scale approximately with $i^2$ but show definite
deviations. Runs with smaller amplitude indicate that this error is
not a nonlinear effect. The error appears to be due to accumulation of
higher order truncation error, mixed with some smaller roundoff error.
Nevertheless, the calculation is still extremely accurate including up
to the end of the run, when deviation from strict second order
convergence is more noticeable.  This is evident from
Fig.~\ref{fig:polyanderr}, which shows the news function and the
$L_{\infty}$ error norm (multiplied by $2000$) for the finest grid.
The error is at least 200 times smaller than the news function; thus
controlled errors of less than one-half of a percent are achievable at
the larger resolutions.

\subsection{Scattering of ingoing pulses}

\begin{figure}
  \centerline{\epsfxsize=3.072in\epsfbox{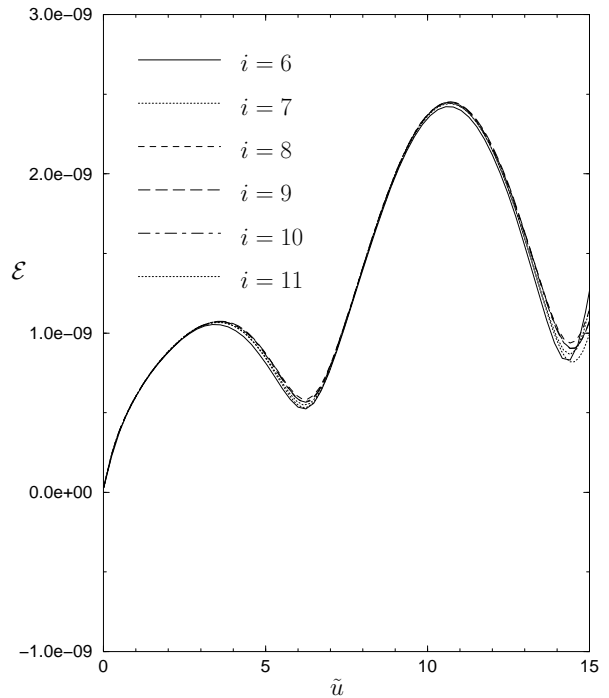}}
  \caption{ The rescaled $L_2$ norm ${\cal E}=(i/6)^2\|
            N_{code} - N_{pert}\|_{2}$ versus grid-parameter $i$ for the
            ingoing pulse test. Second order convergence is confirmed by
            the reasonable overlap of the norms.}
  \label{fig:IngL2}
\end{figure}

These tests correspond to the realistic situation of the scattering of
gravitational radiation which has been created in the near field of a
black hole. We place a compact support pulse on the initial null
hypersurface and set the boundary data on $\hor^-$ to zero. The
initial data were given by Eq.~(\ref{eq:datain}). The setup is
described in Fig.~\ref{fig:setup}.  The perturbation calculation was
carried out using the evolution algorithm for $\hat F_0$ described
in~\cite{close2}.  For this test we choose $x_{min} = 0.6$, $x_{max}
=0.8$, $n=6$, and an amplitude $A$ of $2.6\cdot10^{-6}$.  The
corresponding pulse initially extends from $r=3M$ to $r=8M$.  The
nonlinear runs were performed with the same grid parameters as in
Sec.~\ref{sec:fullpolytest}, with the grid sizes again determined by
the integer $i$ ranging from 6 to 11. The time step was set to $du =
-u\, 5\cdot10^{-4}/ i$, where the factor of $u$ ensures that the CFL
condition remains satisfied (by approximately a factor of 1/4).
Figure~\ref{fig:IngL2} shows the rescaled $L_{2}$ norm of the error
${\cal E}=(i/6)^2 \| N_{code} - N_{pert}\|_{2}$ for the above 6 grids.
The excellent overlap of the norms confirms that the nonlinear null
code is second order convergent.

\begin{figure}
  \centerline{\epsfxsize=3in\epsfbox{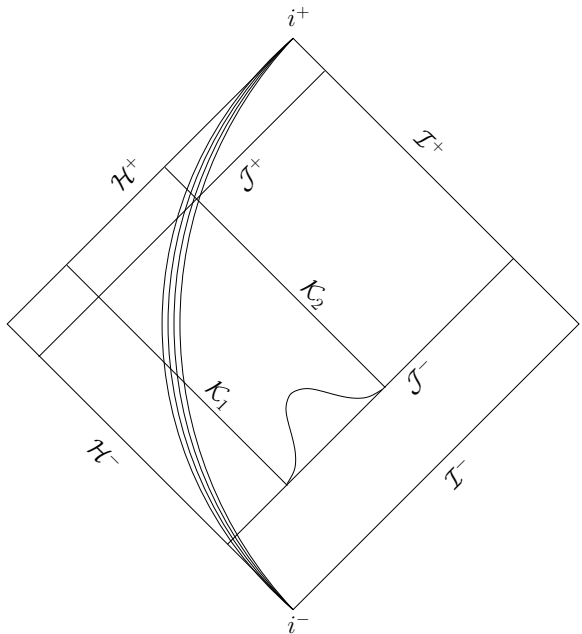}}
  \caption{ The ingoing pulse on a Schwarzschild background.
            The timelike curves (along with $\hor^-$) are the worldlines
            of the first 5 radial gridpoints (the first gridpoint is
            located at $\hor^-$).  The $r=const$ lines begin at past
            timelike infinity $i^-$ and terminate at future timelike
            infinity $i^+$.  An ingoing pulse well resolved on $\JM$ is
            compressed (in the $x$ coordinate) into 3 gridpoints by the
            time it reaches  $\JP$ }
  \label{fig:coord_sing}
\end{figure}

The second order convergence of the news from the ingoing pulse begins
to break down at $\tilde u = 15 M$ as a consequence of our choice of
coordinate system. Fig.~\ref{fig:coord_sing} shows a plot of
Schwarzschild background spacetime with the timelike curves
corresponding to the worldlines of the first few radial gridpoint (the
ingoing null curve $\hor^-$ is the innermost gridpoint). An initially
compact pulse, bounded by the ingoing null lines ${\cal K}_1$ and
${\cal K}_2$, is placed on $\JM$. At the time level $\JP$ the ingoing
pulse will occupy a region containing only a few radial gridpoints.
Consequently, a well resolved pulse on the initial slice will cease to
be resolved in a finite amount of time. Radial resolution of the
ingoing pulse begins to break down at $\tilde u = 15 M$. This
breakdown in resolution happens relatively quickly because we chose a
grid that is uniform in the $x$ coordinate. The evolutions could be
extended by using adaptive mesh refinement techniques recently
developed for characteristic codes \cite{pretoriusCAMR}, and by
different choices for the compact radial coordinate. (Note that the
loss of resolution happens in the radial direction, not in the angular
ones, so that the techniques in~\cite{pretoriusCAMR} could be readily
applied to this case.  We will revisit this point in
Sec.~\ref{sec:concl})

\section{Analysis of mode-mode coupling in gravitational
    radiation scattering}
\label{sec:coupling}

The $(u,x,x^A)$ coordinates, which define the computational grid of
the PITT code~\cite{high,gomez2001} are adapted to the geometry of the
inner boundary. A gravitational wave detector would measure the
radiation as seen in a distant inertial coordinate system, located
essentially at $\scri^+$. We denote by $(u_B,y^A)$ the corresponding
inertial Bondi coordinates on $\scri^+$ (where $x=1$). Thus for the
purposes of gravitational wave data analysis the news function must be
expressed in the form $N(u_B,y^A)$. In the course of this work,
modules have been added to the PITT code to carry out this
transformation, which is essential in order to provide correct time
profiles and harmonic mode analysis. In the linear regime,
$(u,x^A)\approx(u_B,y^A)$ so that these corrections to the news
function are of second order and can be ignored. That is no longer the
case when we consider the nonlinearities manifested by mode coupling.

The ``inertial'' news function $N(u_B,y^A)$ is obtained by performing
a fourth order accurate interpolation between the $(u,x^A)$ and
$(u_B,y^A)$ grids. It is then decomposed into spin-weight 2 spherical
harmonic amplitudes $N_{\ell\,m}$ via second order accurate
integration over the sphere with solid angle $\Omega=4\pi$,
\begin{equation}
     N_{\ell\, m}=\oint N \,{}_2\bar R_{\ell\,m}\, d\Omega.
  \label{eq:Clmdef}
\end{equation}
See App.~\ref{ap:spin} for further details concerning the
spin-weighted harmonic decomposition and how the integration is
carried out on the stereographic patches.

In linear theory, in a gauge where $\beta = O(A^2)$ and $U = O(A)$,
the $N_{\ell\, m}$ coefficients are related to the $J_{\ell\, m}$
coefficients of the metric function $J$  (obtained by applying Eq.
(\ref{eq:Clmdef}) to $J$ rather than $N$) by
\begin{equation}
  N_{\ell\, m} = -\frac{1}{2} r^2 \partial_r \partial_u J_{\ell\, m} -
    \frac{\ell(\ell+1)}{4} \Re{(J_{\ell\, m})}.
\label{eq:NewsCoef}
\end{equation}
In linear theory there is no mode coupling in $J$, and initial data
containing a single ${}_2R_{\ell\,m}$ mode produces a Bondi news
function containing only that mode.

\subsection{Axisymmetric mode coupling} \label{sec:axi}

We first study mode coupling in the axisymmetric case by prescribing
initial data as an ingoing ($\ell=2$, $m=0$) pulse, as in
Eq.~(\ref{eq:datain}) with $n=3$, $x_{min} = .56$, and $x_{max} = .8$.
This choice of initial pulse, extending from $r=2.55M$ to $r=8M$, is
made for computational economy because the nonlinear effects are weak
for $r\gg3M$. We vary the amplitude from $A=2.6\cdot10^{-6}$, well in
the linear regime, to $A=2.6\cdot10^{-1}$, where nonlinear effects can
be clearly observed. We display results obtained with a resolution of
$77\times 77$ angular grid points and $181$ radial grid points. This
grid size (the size of the base grid used in the convergence tests)
allows a broad search for interesting qualitative behavior with
reasonable computational time. The news function is computed in inertial
Bondi coordinates and then decomposed into spin-weight 2 spherical
harmonics.  The reflection symmetry and axisymmetry of the initial data
are preserved by the numerical evolution in the nonlinear regime, so
that the only non-vanishing amplitudes $N_{\ell\, m}$ of the scattered
radiation have even $\ell$ and $m=0$.  In addition, these symmetries
imply that the $N_{\ell\, 0}$ are real, which in our conventions means
that only the $\oplus$ polarization mode is excited.

\begin{figure}
  \centerline{\epsfxsize=3.148in\epsfbox{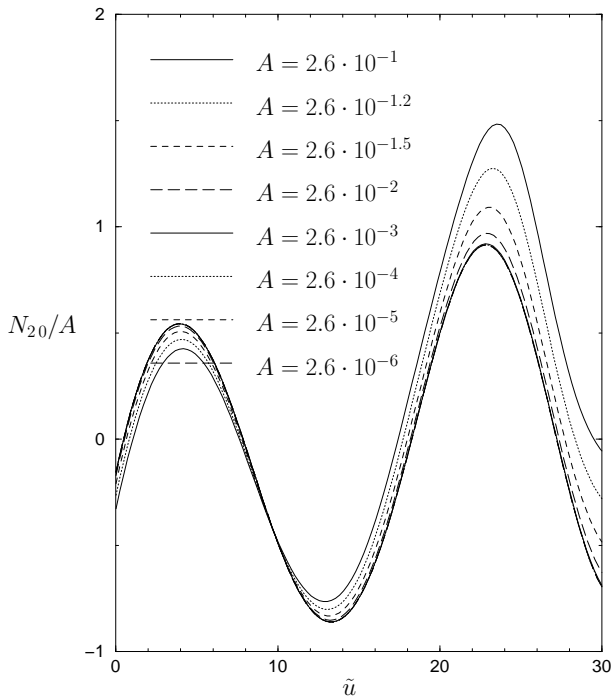}}
  \caption{ The rescaled coefficient $N_{2\,0}/A$ for
            $A=2.6\cdot10^{-1}, 2.6\cdot10^{-1.2}, 2.6\cdot10^{-1.5},
            2.6\cdot10^{-2},\cdots,2.6\cdot10^{-6}$. For
            $A<2.6\cdot10^{-2}$, $N_{2\,0}/A$ has negligible dependence
            on $A$ and the curves overlap.}
  \label{fig:Cl2m0}
\end{figure}

\begin{figure}
  \centerline{\epsfxsize=3.110in\epsfbox{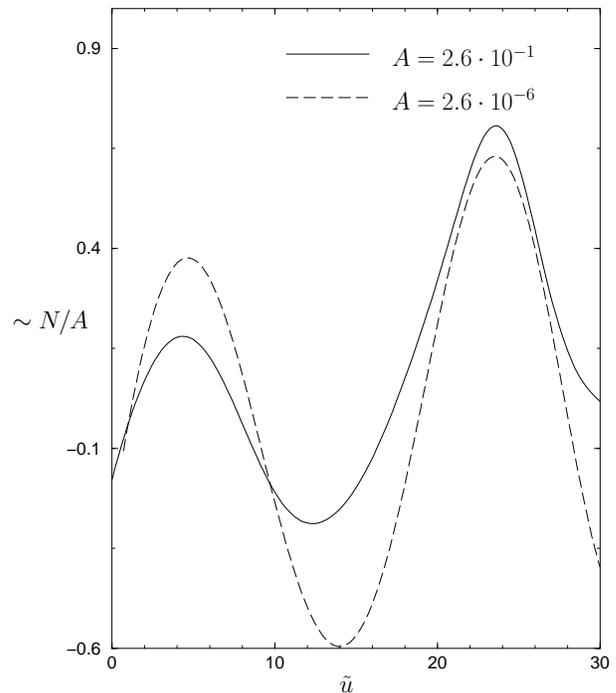}}
  \caption{ The news measured by an observer on the equator
            for $A=2.6\cdot10^{-1}$ and $A=2.6\cdot10^{-6}$. The
            amplitudes have been rescaled by a factor $\sim 1/A$ to make
            comparison of the waveforms easier. }
  \label{fig:news_1_pt_adj}
\end{figure}

The following figures show the time dependence of the lowest order
mode amplitudes, $N_{2\,0}$, $N_{4\,0}$, and $N_{6\,0}$,  which  are
excited in the news function by a given input amplitude $A$.
Figure~\ref{fig:Cl2m0} shows the dependence of $N_{2\,0}/A$ on $A$.
From the figure it is clear that nonlinear effects do not make
$N_{2\,0}$ deviate significantly from its perturbative waveform for
$A<2.6\cdot10^{-2}$. However, at higher amplitudes the effect of
nonlinearity is to amplify the wave, which could enhance the prospects
of detection above the level predicted by linearized theory. This
nonlinear amplification is primarily quadratic in $A$. In addition to
the nonlinear amplification there is also a phase shift which can be
seen more clearly in Fig.~\ref{fig:news_1_pt_adj}.

\begin{figure}
  \centerline{\epsfxsize=3.464in\epsfbox{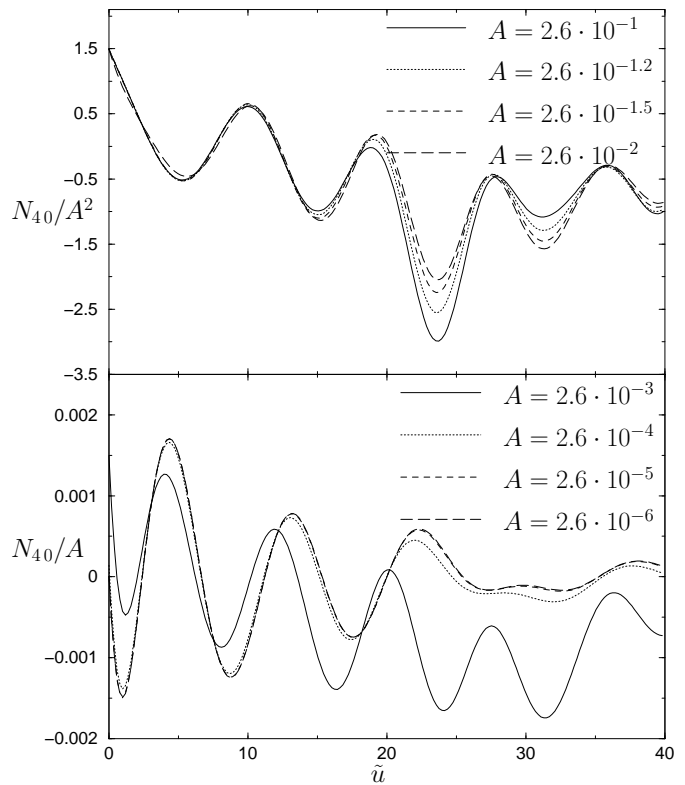}}
  \caption{ The rescaled coefficient $N_{4\,0}/A$ for
            $A=2.6\cdot10^{-3}, 2.6\cdot10^{-4},\cdots,
            2.6\cdot10^{-6}$, and the rescaled coefficient
            $N_{4\,0}/A^2$ for $A=2.6\cdot10^{-1}, 2.6\cdot10^{-1.2},
            2.6\cdot10^{-1.5}, 2.6\cdot10^{-2}$.  Note that at early
            times $N_{4\,0}/A$ is independent of $A$ for
            $A<2.6\cdot10^{-4}$.  For small $A$, the computed value of
            $N_{4\,0}$ is dominated by truncation error and is thus
            proportional to $A$. For larger $A$, where its computed
            value is of physical relevance, $N_{4\,0}$ is proportional
            to $A^2$. When $A=2.6\cdot10^{-3}$, the coefficient contains
            a mixture of order $O(A)$ truncation error and order
            $O(A^2)$ non linear terms. }
  \label{fig:Cl4m0}
\end{figure}

Nonlinear effects are more easily seen in the modes that vanish in the
linearized approximation. Figure~\ref{fig:Cl4m0} shows that $N_{4\,0}$
scales as $A^2$ at sufficiently high amplitudes. For smaller
amplitudes $A\,$ this quadratic scaling is masked by the truncation
error introduced in the computation of the spherical harmonic
decomposition of the news function at the given grid size and, as a
result, $N_{4\,0}$ scales roughly linearly with $A$. 

\begin{figure}
  \centerline{\epsfxsize=3.767in\epsfbox{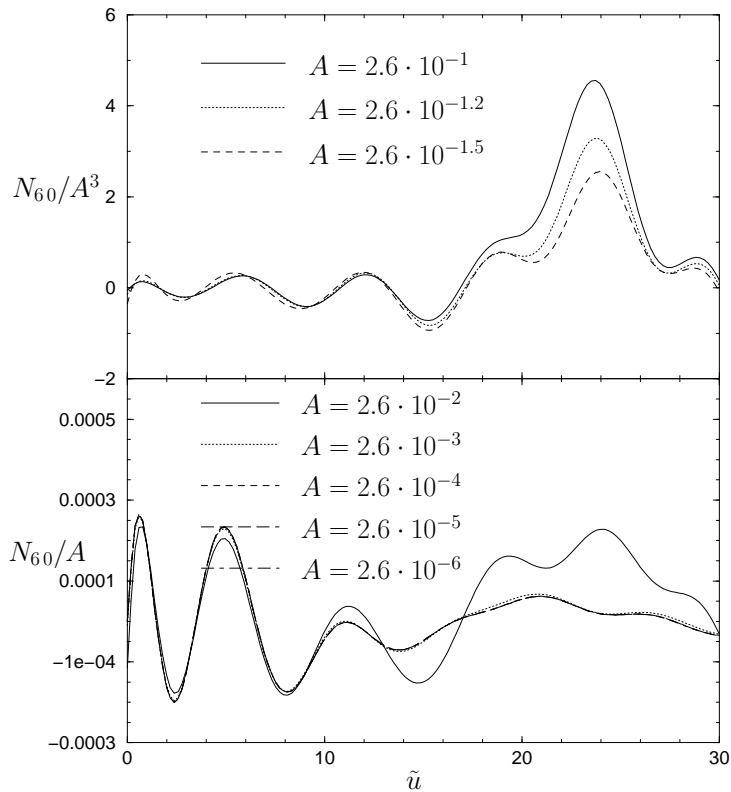}}
  \caption{ The rescaled coefficient $N_{6\,0}/A$ for
            $A=2.6\cdot10^{-2}, \cdots,2.6\cdot10^{-6}$, and the
            rescaled coefficient $N_{6\,0}/A^3$ for $A=2.6\cdot10^{-1},
            2.6\cdot10^{-1.2}, 2.6\cdot10^{-1.5}$.  Note that
            $N_{6\,0}/A$ is independent of $A$ for $A<2.6\cdot10^{-2}$.
            For small $A$, the value of the $N_{6\,0}$ is dominated by
            truncation error and is thus proportional to $A$.  For
            larger $A$, the coefficient is proportional to $A^3$.
            However, the nonlinear terms in $N_{6\,0}$ are smaller than
            truncation error for $A$ as large as $2.6\cdot10^{-2}$.  }
  \label{fig:Cl6m0}
\end{figure}

Figure~\ref{fig:Cl6m0} shows $N_{6\,0}$ scales as $A^3$ at the highest
amplitudes. As might be expected for a higher $\ell$ mode, the masking
effects of truncation error are now more severe at the lower
amplitudes.  The breakdown of scaling behavior in
Fig's.~\ref{fig:Cl2m0} - \ref{fig:Cl6m0} results from inaccuracy at
times beyond $\tilde u = 20$ due to lack of resolution of short scale
features using the current grid. These short scale features arise
inside $r=3M$ where quasinormal ringing is known to originate.  

\begin{figure}
  \centerline{\epsfxsize=2.772in\epsfbox{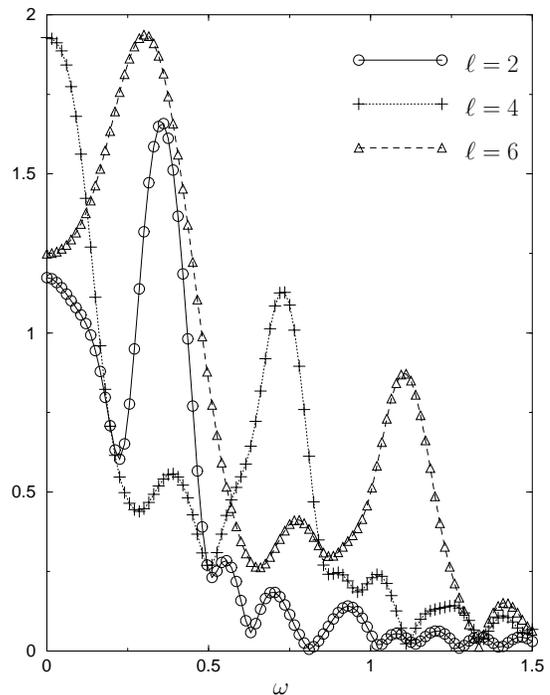}}
  \caption{ Fourier transforms (after rescaling to similar
            amplitudes to facilitate comparison of peak locations) of
            the $N_{2\,0}$, $N_{4\,0}$, and $N_{6\,0}$ time profiles for
            the $A=2.6\cdot10^{-1}$ run.  The ($\ell=2$, $m=0$) mode
            shows a strong peak at $\omega \approx .36$, the ($\ell=4$,
            $m=0$) mode shows a strong peak at $\omega \approx .73$ and
            $\omega\approx 0$, and the ($\ell=6$, $m=0$) mode shows a
            strong peak at $\omega \approx 1.10$ and $\omega\approx
            .30$.}
  \label{fig:four}
\end{figure}

There are some understandable aspects in these mode coupling results.
Apart from numerical truncation error effects at low amplitudes, we
find that the corresponding amplitudes $N_{\ell\,0}$ scale as
$A^{\ell/2}$. This is consistent with the property of ordinary
spherical harmonics
$$
\left(Y_{\ell\, m}\right)^j  \sim Y_{(j\ell)\,m} +
 \mbox{terms with smaller $\ell$}.
$$
Thus we expect that an $\ell = 2 j$ mode arises from order $j$ (and
higher) nonlinear terms, and hence will scale as $A^j + O(A^{j+1})$.
The theory of how the Einstein equations couple these spin-weighted
spherical harmonics has not been worked out. A full analysis of the
possible modes produced at a given order would require a computation
of the appropriate Clebsch-Gordon coefficients for spin-weighted
spherical harmonics. Finding these coefficients is complicated by the
fact that there are many ways to combine $\ell$ and $m$ modes of
various spin-weights  to produce a spin 2 function (e.g.
${}_4R_{\ell\,m}\,\,{}_2\bar R_{\ell'\,m'}$; ${}_1R_{\ell\,m}\,\,
{}_1R_{\ell'\,m'}$; ${}_0R_{\ell\,m}\,\,{}_2R_{\ell'\,m'}$; $\cdots$).
Rather than work out these  coefficients we use a more naive approach
of looking at the possible modes produced by combinations of
spin-weight 0  harmonics. From this approach we find that the possible
modes produced by combining an ($\ell=2$, $m=0$) mode with itself
(i.e. quadratic coupling) are ($\ell=4$, $m=0$) and ($\ell=2$, $m=0$)
[$\ell=0$ is not allowed  for spin 2 fields]. Thus the ($\ell=2$,
$m=0$) mode in the news would contain linear and higher order terms,
whereas the ($\ell=4$, $m=0$) mode would contain quadratic and higher
terms.   The above numerical results are consistent with these naive
expectations.  In addition, there is a similar behavior in the
frequencies of the various $\ell$ modes. Figure \ref{fig:four} shows
the Fourier transforms of the  $N_{2\,0}$, $N_{4\,0}$, and $N_{6\,0}$
coefficients for the $A=2.6\cdot10^{-1}$ run. The  ($\ell=2$, $m=0$)
mode shows a strong peak with maximum at $\omega \approx .36$; the
($\ell=4$, $m=0$) mode has peaks at $\omega \approx .73$ and $\omega
\approx 0$; and the ($\ell=6$, $m=0$) mode has peaks at $\omega
\approx 1.10$ and $\omega \approx .30$. This entire behavior is
exactly what would arise from the nonlinear power law response
$f^{\ell/2}$ to an $\ell=2$ mode $f$; i.e.\ quadratic terms produced
from $f\sim \sin\omega_0 t$ contains frequencies $2 \omega_0$ and $0$.
Similarly, cubic terms produce frequencies $\omega_0$ and $3
\omega_0$.

The dominant contribution to the frequency of the ($\ell=2$, $m=0$)
mode is expected to be the lowest $\ell=2$ quasinormal frequency. For
an $M=1$ black hole the lowest $\ell=2$ quasinormal frequency is
$\omega = .373672$ \cite{nollert}.  However, the $A=2.6\cdot10^{-1}$
amplitude pulse should contribute significant mass to the system and a
lower frequency should be expected, in agreement with the measured
peak at $\omega \approx .36$. To measure the accuracy of the
frequencies obtained for these systems we performed a similar analysis
with the $A=2.6\cdot10^{-6}$ run. In this low amplitude limit, the
Fourier transform of $N_{2\,0}$ showed a maximum at $\omega \approx
.37$, very close to the frequency of the lowest quasinormal mode. 

\begin{figure}
  \centerline{\epsfxsize=3.110in\epsfbox{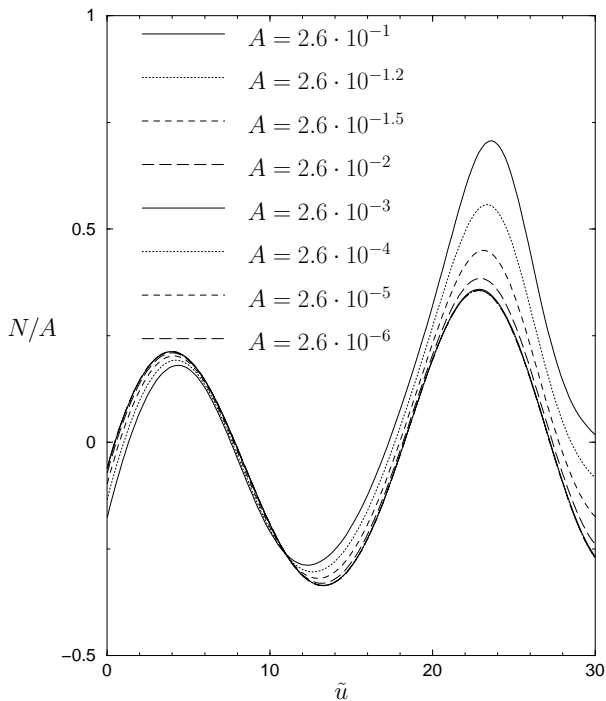}}
  \caption{ The rescaled Bondi news function $N/A$ on the
            equator. The plot is very similar to Fig.~\ref{fig:Cl2m0}
            (the plots differ by an overall factor which results from
            the normalization of ${}_2R_{2\,0}$ and phase shifting in
            the larger amplitude runs) due to the dominance of the
            ($\ell=2$, $m=0$) mode. The news is axisymmetric and
            contains only the $\oplus$ mode. }
  \label{fig:news_1pt}
\end{figure}

Of direct importance for designing templates for wave detection is the
waveform obtained by the net superposition of these modes. In
Fig.~\ref{fig:news_1pt} we plot the time dependence of the inertial
news function, as measured by an observer at the equator in a
coordinate system adapted to $\scri^+$, as a function of input
amplitude. The figure shows nonlinear amplification of the news
function with increasing input amplitude.  The peaks of the waveform
differ by up to a factor of 2 from the values which would be obtained
by a linearized calculation. 

In addition there are phase shifts in the location of the peaks. These
are apparent in Fig.~\ref{fig:news_1_pt_adj} which compares $N/A$ at
the equator for the nonlinear case $A=2.6\cdot10^{-1}$ and the linear
case $A=2.6\cdot10^{-6}$. The difference between these two waveforms
corresponds to the error that would be made by using a linearized
calculation to obtain the $A=2.6\cdot10^{-1}$ waveform. In the first
oscillation the $A=2.6\cdot10^{-1}$ run has the larger wavelength,
while afterward its wavelength is shorter.

Accurate knowledge of the phase is very important to data analysis for
extracting the signal~\cite{Cutler:1993tc}. When using matched
filtering over a number of cycles of the waveform, the total
integrated error in the phase must be no greater than 10\% of a cycle.
For example, a phase shift of 2\% per cycle from the value programmed
into the detection template would render the phase information useless
in 5 cycles. The trend displayed in Fig's.~\ref{fig:news_1pt} and
\ref{fig:news_1_pt_adj} is a drift in phase of about 15\% from the
first maximum to the first minimum. The nonlinear news oscillates
quicker in this  first half cycle. This trend is  reversed in the
second half of the first cycle and the net phase shift relative to the
linear news after the first complete cycle is $2\%$. Beyond the first
cycle the nonlinear news exhibits  a consistently shorter oscillation
period than the linear news. These results are of potential importance
but, as already cautioned, at the current resolution it is not clear
how accurate they are past $\tilde u \approx 20$.


\subsection{Azimuthal mode coupling} \label{sec:nonaxi}

We study azimuthal effects of mode coupling by prescribing initial
data as an ingoing ${}_2R_{2\,2}$ pulse. Again we give the data in the
form of Eq.~(\ref{eq:datain}) with $n=3$, $x_{min} = .56$, and
$x_{max} = .8$, we vary the amplitude from $A=10^{-6}$ to $A=.36$ and
we use a fixed grid of $77\times 77$ angular grid points and $181$
radial grid points (larger grids are used when analyzing the cubic
modes). The parity and reflection symmetries of the initial data are
preserved by the nonlinear evolution, so that the only non-vanishing
amplitudes $N_{\ell\, m}$ of the scattered radiation have even values
of $\ell$ and even $m \geq0$.  However, in this case, both the
$\oplus$ and $\otimes$ polarization modes are excited. We decompose
the resulting news function in terms of the spin-weighted functions
${}_2R_{\ell\,m}$ up to $\ell=6$. The non-vanishing quadratic modes
present in the news were the ${}_2R_{2\,0}$, ${}_2R_{4\,0}$, and
${}_2R_{4\,4}$ harmonics. The non-vanishing cubic modes were the
${}_2R_{4\,2}$, ${}_2R_{6\,2}$ and ${}_2R_{6\,6}$ harmonics.
 
\begin{figure}
  \centerline{\epsfxsize=3.248in\epsfbox{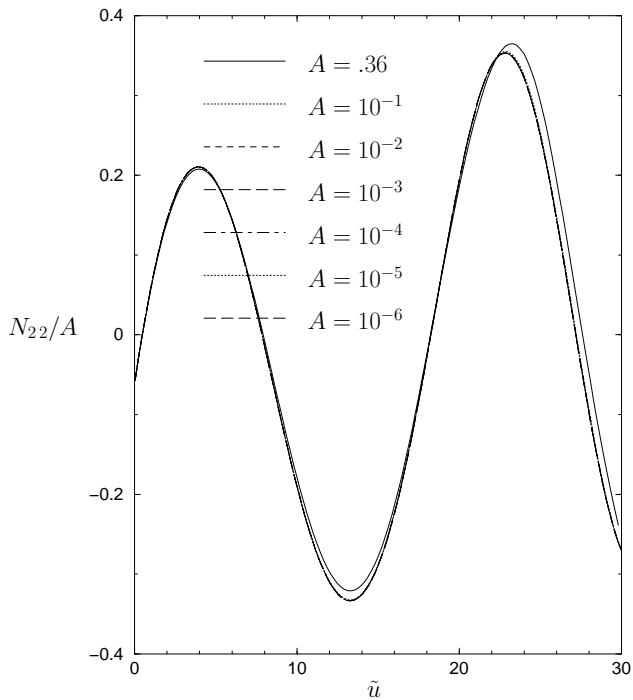}}
  \caption{ Azimuthal coupling. The rescaled coefficient
            $N_{2\,2}/A$ is plotted versus time.  Note the near perfect
            linear dependence of $N_{2\,2}$ on $A$ for $A\le.1$.
            Nonlinear effects are only evident for the largest
            amplitude.}
  \label{fig:nsr_22}
\end{figure}

Fig.~\ref{fig:nsr_22} shows the coefficient $N_{2\,2}/A$ versus $A$.
In linear theory the dependence of $N_{2\,2}$ on $A$ is linear, and
this is observed for the entire run when $A\le.1$. Nonlinear
amplification and phase shifts are only apparent for $A=.36$. 

\begin{figure}
  \centerline{\epsfxsize=3.248in\epsfbox{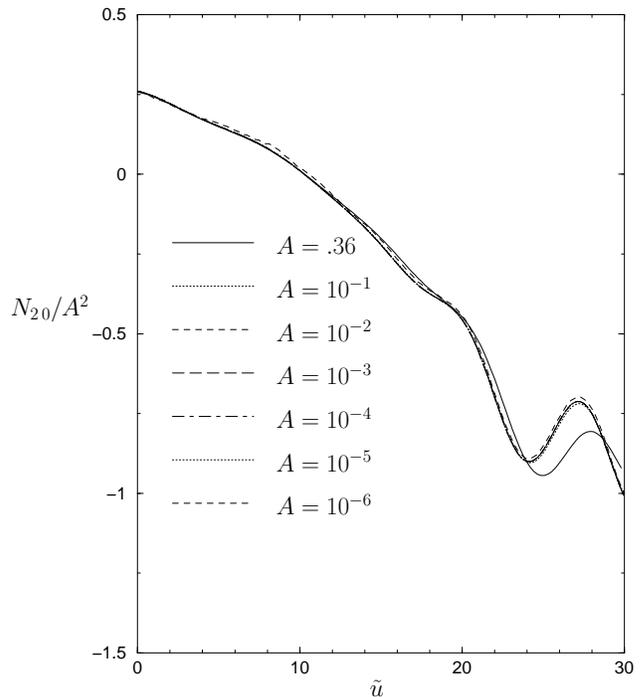}}
  \caption{ Azimuthal coupling. The rescaled coefficient
            $N_{2\,0}/A^2$ is plotted versus time.  Note the near
            perfect quadratic dependence of $N_{2\,0}$ on $A$ for $A\le
            .1$.  Hidden  within the low frequency signal at $\tilde u <
            15$ there is a higher frequency mode as is apparent in
            Fig.~\ref{fig:nsr_four}.  The slight deviation of the
            $A=10^{-6}$ curve is due to roundoff error. }
  \label{fig:nsr_20}
\end{figure}

\begin{figure}
  \centerline{\epsfxsize=3.248in\epsfbox{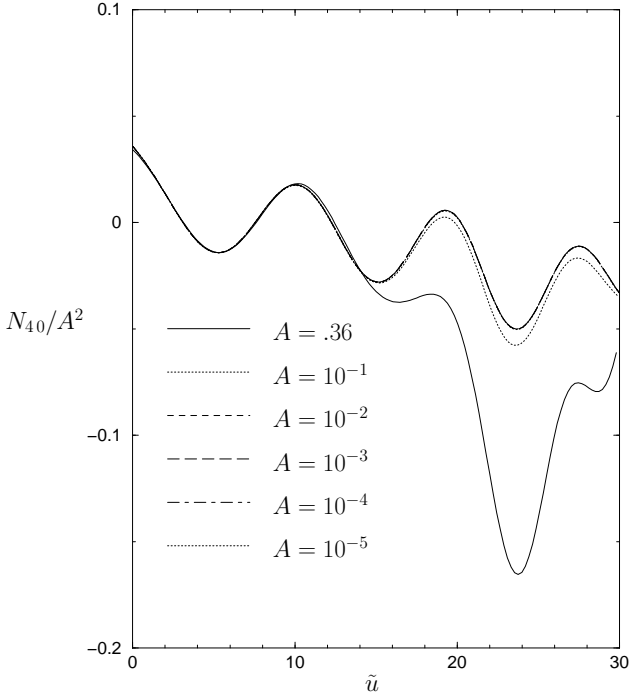}}
  \caption{ Azimuthal coupling: The rescaled coefficient
            $N_{4\,0}/A^2$ is plotted versus time.  Note the near
            perfect quadratic dependence of $N_{4\,0}$ on $A$ at early
            times.  The late-time deviation of the $A=.36$ curve may be
            due to the loss of resolution near $r=2M$.  The frequency is
            roughly twice that of $N_{2\, 2}$.}
  \label{fig:nsr_40}
\end{figure}

\begin{figure}
  \centerline{\epsfxsize=3.248in\epsfbox{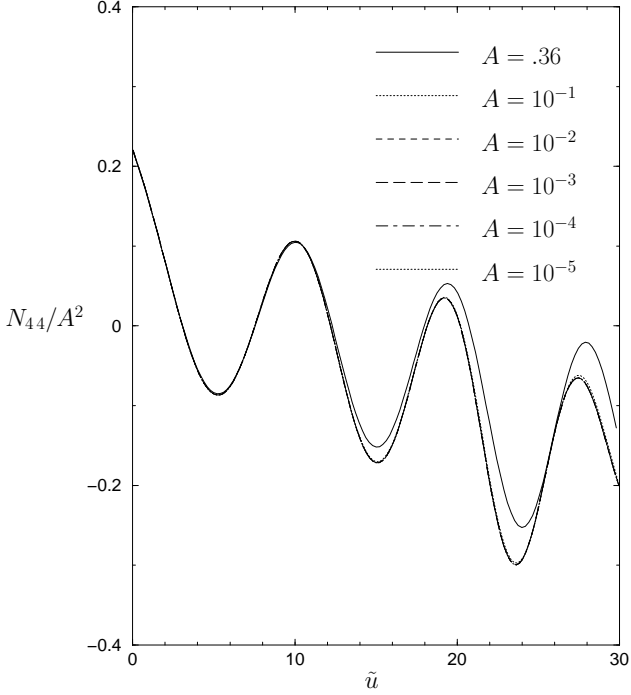}}
  \caption{ Azimuthal coupling: The rescaled coefficient
            $N_{4\,4}/A^2$ is plotted versus time.  Note the near
            perfect quadratic dependence of $N_{4\,4}$ on $A$ for
            $A\le.1$.  The late time deviation is only significant for
            $A \geq .1$.  The frequency is roughly twice that of $N_{2\,
            2}$.}
  \label{fig:nsr_44}
\end{figure}

Fig's.~\ref{fig:nsr_20},~\ref{fig:nsr_40},~and~\ref{fig:nsr_44} graph
the coefficients $N_{2\,0}/A^2$, $N_{4\,0}/A^2$, and $N_{4\,4}/A^2$
respectively.  In each case there is a near perfect early time overlap
between the curves at different amplitudes for $A\le.1$.  These
nonlinear modes exhibit a quadratic dependence on $A$ except for a
late time amplification of the $A=.36$ curves in all modes and a late
time amplification of the  $A=.1$ curve of the $N_{4\,0}$ mode. This
late time amplification results from cubic and higher terms
contributing to the modes. The very strong  late-time behavior
observed in $N_{4\,0}$ may result from the loss of  resolution near
$r=2 M$.

\begin{figure}
  \centerline{\epsfxsize=3.248in\epsfbox{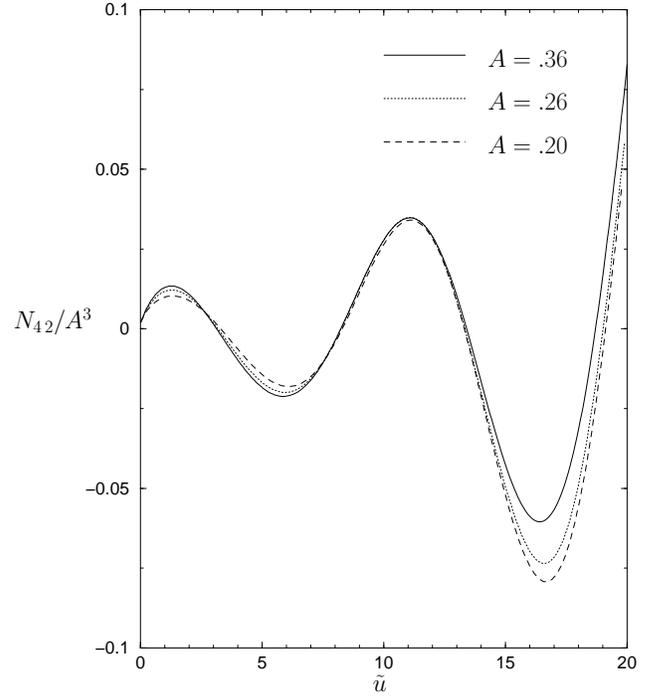}}
  \caption{ Azimuthal coupling: The rescaled coefficient
            $N_{4\,2}/A^3$ is plotted versus time.  The overlap of the
            curves is reasonably good, but the frequency does not
            correspond to the expected behavior for a cubic mode. }
  \label{fig:nsr_42}
\end{figure}

\begin{figure}
  \centerline{\epsfxsize=3.248in\epsfbox{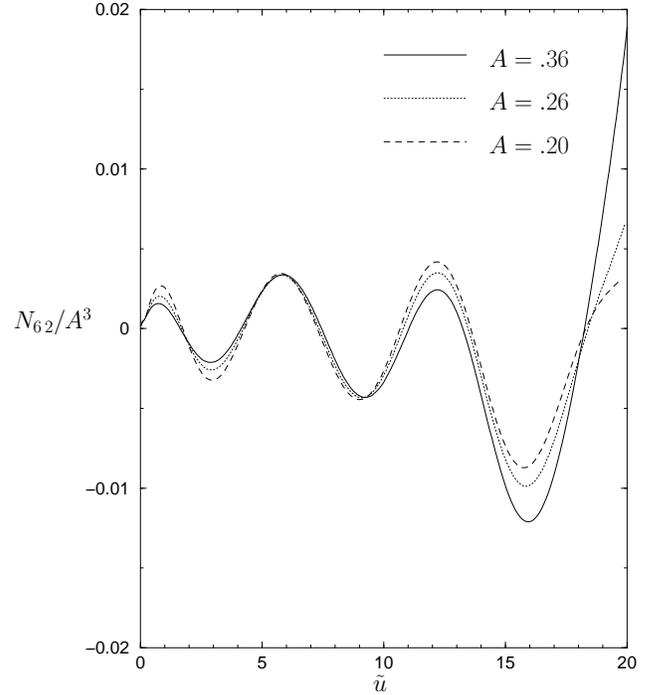}}
  \caption{ Azimuthal coupling: The rescaled coefficient
            $N_{6\,2}/A^3$ is plotted versus time.  The overlap of the
            curves is reasonably good. Tests with higher resolution
            ($269\times269$ angular gridpoints) indicate that the poor
            early time scaling is due to truncation error. The frequency
            is roughly three times that of the $N_{2\,2}$ mode. }
  \label{fig:nsr_62}
\end{figure}

\begin{figure}
  \centerline{\epsfxsize=3.119in\epsfbox{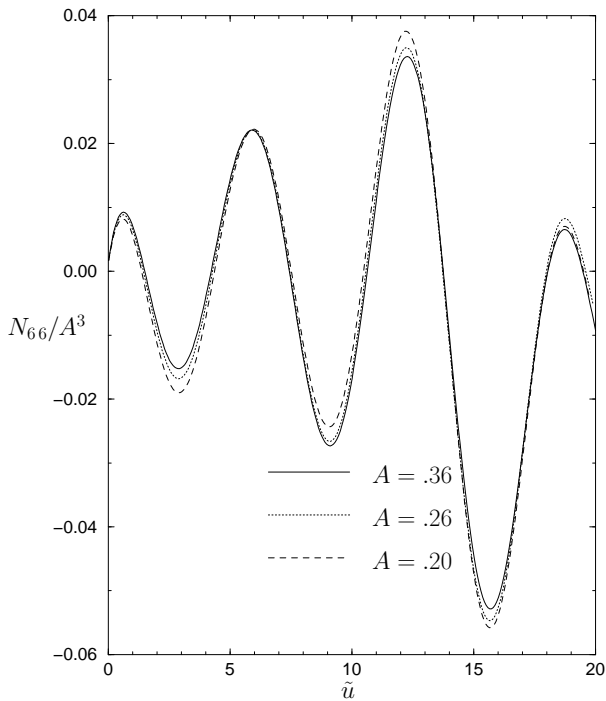}}
  \caption{ Azimuthal coupling: The rescaled coefficient
            $N_{6\,6}/A^3$ is plotted versus time.  The overlap of the
            curves is reasonably good. The frequency is roughly three
            times that of the $N_{2\,2}$ mode. }
  \label{fig:nsr_66}
\end{figure}

Figures~\ref{fig:nsr_42},~\ref{fig:nsr_62},~and~\ref{fig:nsr_66} graph
the coefficients $N_{4\,2}/A^3$, $N_{6\,2}/A^3$, and $N_{6\,6}/A^3$.
These modes show a cubic dependence on $A$, and were obtained by using
a computational grid of $137\times137$ angular gridpoints and $331$
radial gridpoints (roughly twice the resolution of the previous runs).
The amplitudes for these runs were $A=.20$, $A=.26$, and $A=.36$. 

\begin{figure}
  \centerline{\epsfxsize=2.779in\epsfbox{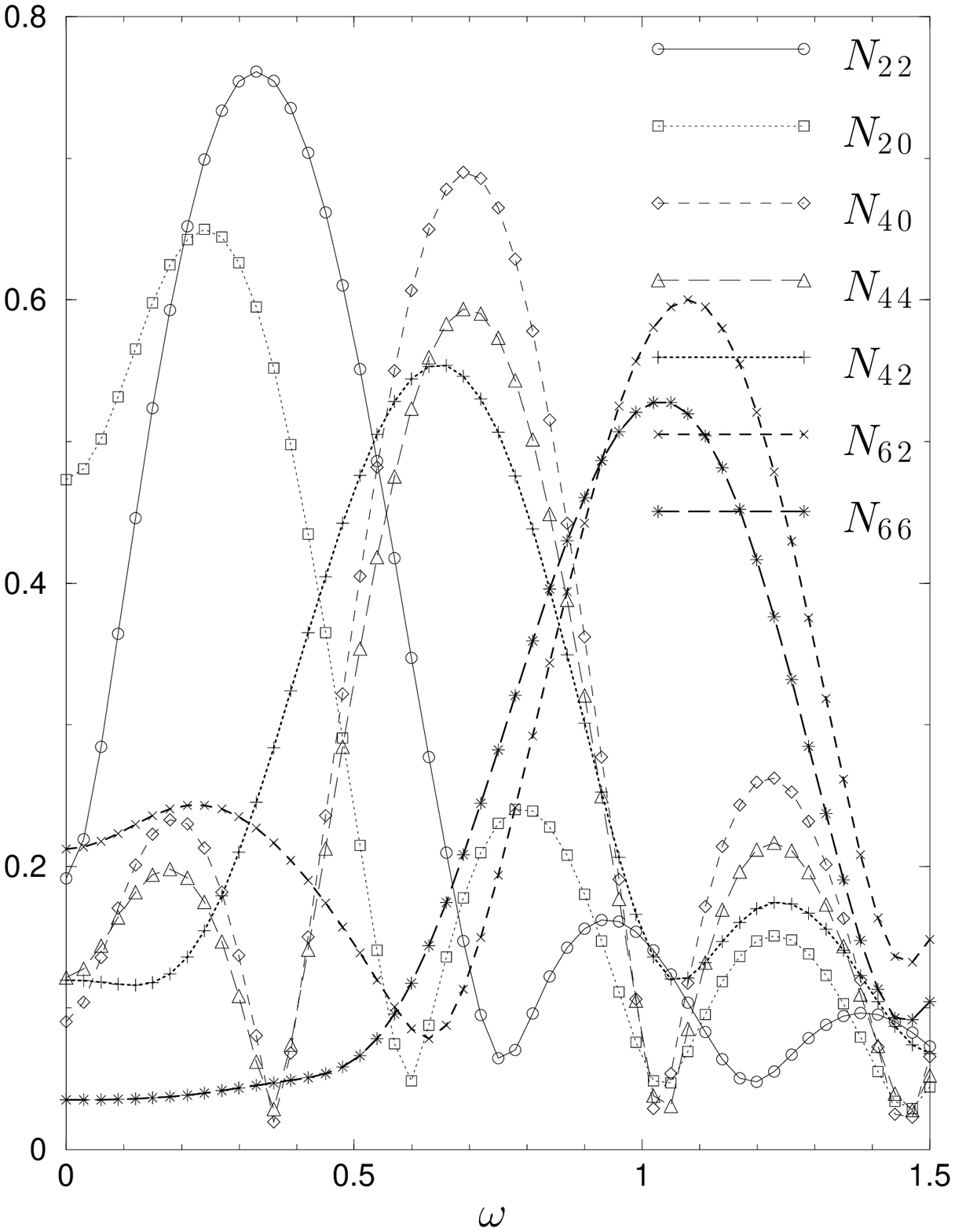}}
  \caption{ Azimuthal coupling, The Fourier response
            (absolute value) to an  $A=.36$ pulse is shown for the first
            few modes. The Fourier integrals were restricted to $\tilde
            u < 15$ and the resulting functions were rescaled to all
            have similar sizes. The dominant frequency of the $N_{2\,2}$
            mode is $\omega \approx .33$.  The $N_{2\,0}$ mode shows a
            strong low frequency peak at $\omega \approx .18$ and a
            weaker peak at $\omega \approx .80$. Interestingly, the
            transforms of $N_{4\,0}$ and $N_{4\,4}$ are very similar
            with both having strong peaks at $\omega \approx .70$. The
            $N_{4\,2}$ mode (a cubic mode) shows an unexpected strong
            peak at $\omega \approx .63$.  The $N_{6\,2}$ mode shows a
            strong peak at $\omega \approx 1.23$ while the $N_{6\,6}$
            mode has a strong peak at $\omega\approx 1.02$. }
  \label{fig:nsr_four}
\end{figure}

The early time behavior of the ($\ell=2$, $m=0$) mode, unlike the
behavior of the other nonlinear modes, is dominated by a low frequency
component. However, hidden in this large signal is a higher frequency
signal with roughly twice the frequency of the ($\ell=2$, $m=2$) mode.
Fig.~\ref{fig:nsr_four} shows the Fourier decomposition for these
modes. The ($\ell=2$, $m=2$) mode has a strong peak at $\omega \approx
.33$, the ($\ell=2$, $m=0$) mode has  strong peaks at $\omega \approx
.80$ and $\omega\approx .18$,  and both the ($\ell=4$, $m=0$) and the
($\ell=4$, $m=4$) modes have strong peaks at $\omega \approx.70$.
Given the large widths of the peaks, they are roughly consistent with
the expected behavior of modes generated by quadratic response to an
($\ell=2$, $m=0$) mode.  On the other hand the frequency spectrum of
the ($\ell=4$, $m=2$) mode has a maximum at $\omega \approx .63$. The
expected frequency of a cubic mode is either the  frequency of the
principal mode or three times the principal frequency. This drift
toward larger than expected frequencies mirrors the behavior of the
($\ell=2$, $m=0$) mode. The ($\ell=6$, $m=2$) mode has a maximum at
$\omega \approx 1.23$ while the  ($\ell=6$, $m=6$) mode has a maximum
at $\omega\approx 1.02$. These latter two results are roughly
consistent with the expected behavior for cubic modes. 


\begin{figure}
  \centerline{\epsfxsize=3.1in\epsfbox{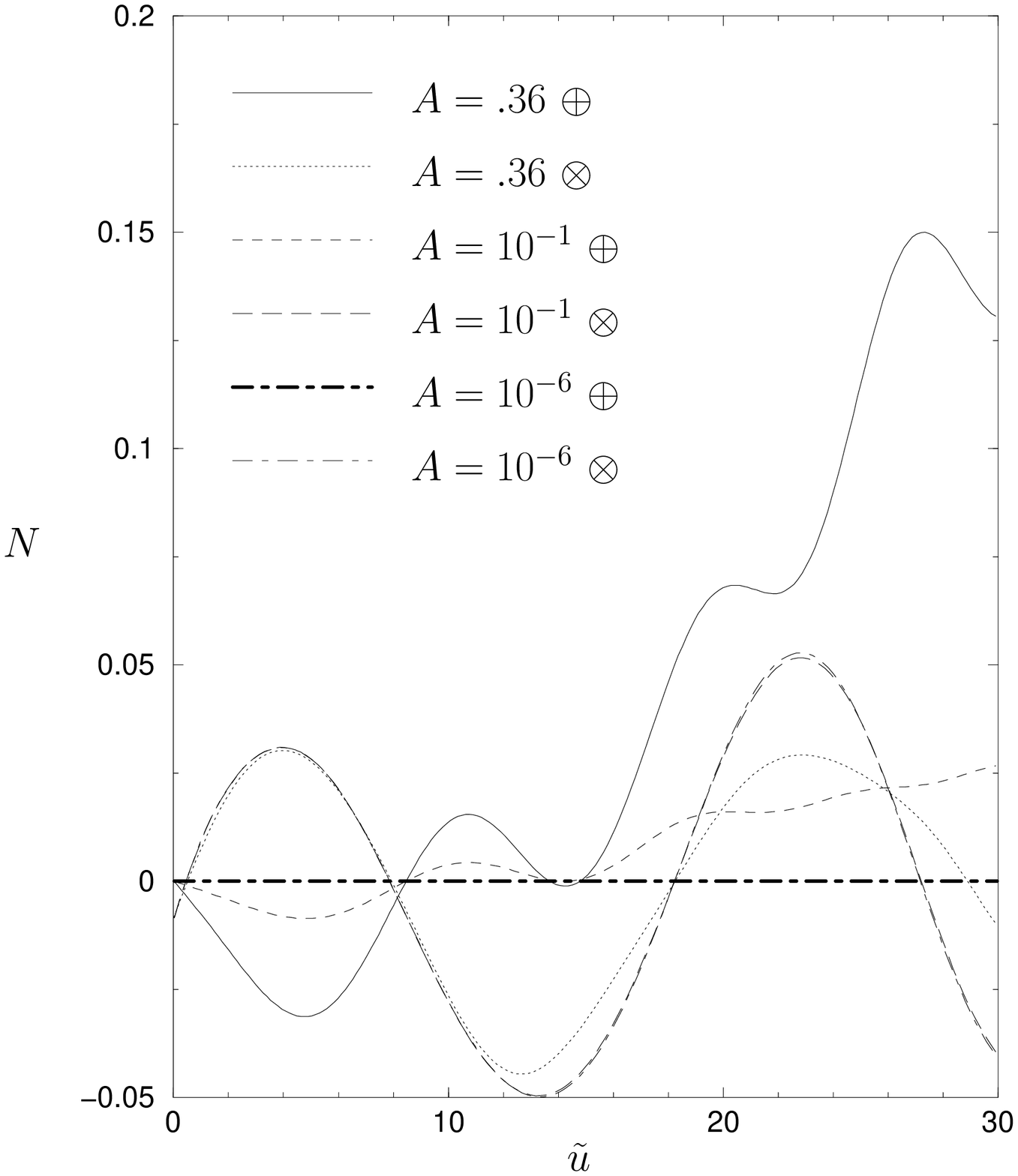}}
  \caption{ Azimuthal $\otimes$, $\oplus$ coupling. The news
            is shown in the $q=p=.5$ observation direction and has been
            rescaled by $1/A$. The rescaled $\otimes$ component of the
            $A=10^{-1}$ and $A=10^{-6}$ runs overlap. Nonlinear changes
            in amplitude and phase are visible in the $\otimes$
            component of the $A=.36$ run. }
  \label{fig:nsr_couple}
\end{figure}

Fig.~\ref{fig:nsr_couple} shows the Bondi news observed at $q=p=.5$
($\theta\simeq 70^o, \phi=45^o$) for the $A=.36$, $A=10^{-1}$, and
$A=10^{-6}$ non-axisymmetric runs.  In the linearized approximation,
the news in this angular direction is always imaginary, corresponding
to pure $\otimes$ polarization.  Nonlinear effects couple the $\oplus$
and $\otimes$ modes.  Initial data with $A=.1$  produce a significant
$\oplus$ component with amplitude roughly 28\% of the $\otimes$
component; while  initial data with $A=.36$ (at $\tilde u = 20$)
produces an $\oplus$ component 28\%  {\em larger} than the $\otimes$
component. Note that there is no significant nonlinear change in
amplitude for the $\otimes$ component for $A\leq .1$. However, if a
gravitational wave detector were not precisely oriented to measure the
$\otimes$ component, significant nonlinear amplification and phase
shifting would arise from the superposition of the $\oplus$ and
$\otimes$ components. In the axisymmetric case, all of the
gravitational radiation is in the $\oplus$ mode, and a gravitational
wave detector would see the same nonlinear amplification and phase
shifts regardless of orientation. In the present case the degree of
nonlinear amplification and phase shifting depends on the orientation
of the detector. 

We can gain insight into this problem by looking at the behavior of
the real, spin-weight zero spherical harmonics $R_{\ell\,
m}={}_0R_{\ell\,m}$ (in a similar way as done in Sec.~\ref{sec:axi}).
For $m$ even and positive, $R_{\ell\, m}$ is reflection symmetric
about the $q$ axis and has even parity with respect to the $(q,p)$
coordinate system. For $m$ even and negative, $R_{\ell\, m}$ has even
parity but is reflection antisymmetric about the $q$ axis. Any product
of positive, even $m$ harmonics will contain these two symmetries and
is therefore a sum of even, positive $m$, real spherical harmonics.
Along with the reflection symmetry of Einstein's equations, this would
seem to imply that data consisting of a positive, even $m$ harmonic
with even $\ell$ would yield a Bondi news function consisting solely
of positive, even $m$ and even $\ell$ harmonics. In addition, we can
predict the possible modes generated by an ($\ell=2$, $m=2$) pulse and
in what order they appear.  For example, the quadratic modes produced
by ($\ell=2$, $m=2$) initial data are ($\ell=2$, $m=0$), ($\ell=4$,
$m=0$), and ($\ell=4$, $m=4$). Note that there is no quadratic
($\ell=2$, $m=2$) response. Thus quadratic terms in the Einstein
equation do not affect the principal mode. Consequently there is no
apparent nonlinear amplification or phase shift in this mode for
$A\leq.1$ (see Fig.~\ref{fig:nsr_22}). Cubic terms can generate an
($\ell=2$, $m=2$) mode so we expect those terms to produce nonlinear
amplification and phase shifts when they become significant.

\subsection{Generating catalogs of waveforms}

The results presented in Sec's.~\ref{sec:axi} and \ref{sec:nonaxi}
suggest a method for efficiently producing a catalog of waveforms
produced  by a given profile of the initial data with amplitude
smaller than, say, $A=.1$.  First perform a run with the linearized
code to generate the linear part of the waveform. Then perform a
single nonlinear run at an amplitude of $A=.1$. One then subtracts off
the linear contribution to the various ($\ell$, $m$) modes and obtains
the quadratic and cubic terms. The scaling with amplitude $A$ of each
of the ($\ell$, $m$) modes can be determined by the arguments given in
Sec's.~\ref{sec:axi}~and~\ref{sec:nonaxi}, and the net waveform for
any $A\lesssim.1$ can be  reconstructed.

For example, consider the waveform generated from an axisymmetric
($\ell=2$, $m=0$) input pulse with an amplitude $A=.1$. We first
subtract off the linear part of the ($\ell=2$, $m=0$) profile to get
the quadratic part of the ($\ell=2$, $m=0$) mode.  We then assume a
scaling of $A^2$ for this part of the ($\ell=2$, $m=0$) mode and the
($\ell=4$, $m=0$) mode, as well as a cubic scaling for the  ($\ell=6$,
$m=0$) mode. We  can then produce the news for all $A\lesssim.1$ from
a single nonlinear run. 

\section{Conclusion}
\label{sec:concl}

We have shown how a characteristic code can be employed to investigate
the nonlinear response of a black hole to the infall of gravitational
wave energy.  This problem, which is of importance to the observation
and interpretation of gravitational waves, can in this way be studied
with a mature, reliable evolution code. In this paper we have focused
on the onset of nonlinear behavior in the waveform produced by the
scattering of a pulse of radiation incident on a Schwarzschild black
hole. However, given sufficient resolution, the same computational
approach could be extended to many other scenarios, such as the
waveform emitted by the dispersion of a pulse of radiation propagating
approximately along the $r=3M$ (unstable) orbit about a Schwarzschild
black hole, as we might expect from the gravitational perturbation
associated with a bounded distribution of matter in such an orbit. 

Besides the computation of accurate waveforms, our study reveals
several features of qualitative importance:

\begin{itemize}

\item {\bf I}. The mode coupling amplitudes consistently scale as
powers $A^n$ of the input amplitude $A$ corresponding to the nonlinear
order of the terms in the evolution equations which produce the mode.
This allows much economy in producing a waveform catalog. Given the
order $n$ associated with a given mode generation, the response to any
input amplitude $A$ can be obtained from the response to a single
reference amplitude $A_0$.

\item {\bf II}. The frequency response has similar behavior but in a
less consistent way. The dominant frequencies produced by mode
coupling are in the approximate range of the quasinormal frequency of
the input mode and the expected sums and difference frequencies
generated by the order of nonlinearity.

\item {\bf III} Large phase shifts, ranging up 15\% in a half cycle
relative to the linearized waveform, are exhibited in the news
function obtained by the superposition of all output modes, i.e.  in
the  waveform of observational significance. These phase shifts, which
are important for design of signal extraction templates, arise in an
erratic way from superposing modes with different oscillation
frequencies. This furnishes another strong argument for going beyond
the linearized approximation in designing a waveform catalog for
signal extraction. 

\item {\bf IV} Besides the nonlinear generation of harmonic modes
absent in the initial data, there is also a stronger than linear
generation of gravitational wave output. This provides a potential
mechanism for enhancing the strength of the gravitational radiation
produced during, say, the merger phase of a binary inspiral above the
strength predicted in linearized theory.

\item {\bf V} In the non-axisymmetric case, there is also considerable
generation of radiation in polarization states not present in the
linearized approximation. In our simulations, input amplitudes in the
range  $A=.1$ to $A=.36$ lead to nonlinear generation of a $\oplus$
component which is of the same order of magnitude as the $\otimes$
component (which would be the sole component according to linearized
theory). As a result, significant nonlinear amplification and phase
shifting of the waveform can be observed  depending on the orientation
of a gravitational wave detector. 

\end{itemize}

As already noted by Papadopoulos in his work on axisymmetric mode
coupling~\cite{papad}, these effects arise from three types of
nonlinearity: (i) Modification of the light cone structure governing
the principal part of the equations and hence the propagation of
signals; (ii) Modulation of the Schwarzschild potential by the
introduction of an angular dependent ``mass aspect''; and (iii)
Quadratic and higher order terms in the evolution equations which
couple  the modes. 

Although Papadopoulos studied nonlinear mode generation produced by an
outgoing pulse, as opposed to the case of an ingoing pulse studied
here, these same factors are in play and it is not surprising that
both studies have common features. In both cases, the major nonlinear
effects arise in the region near $r=3M$. Analogs of items {\bf I} -
{\bf IV} above are all apparent in Papadopoulos's results.  At the
finite difference level, both codes respect the reflection symmetry
inherent in Einstein's equations and exhibit the corresponding
selection rules arising from parity considerations. In the
axisymmetric case considered by Papadopoulos, this forbids the
nonlinear generation of a $\oplus$ mode  from a $\otimes$ mode, as in
item {\bf V} above.

The evolution along ingoing null hypersurfaces in the work of
Papadopoulos  and the evolution along outgoing null hypersurfaces in
the present work have complementary numerical features. The grid based
upon {\em ingoing} null hypersurfaces avoids the difficulty depicted
in Fig. \ref{fig:coord_sing} in resolving effects close to $r=2M$ with
a grid based upon {\em outgoing} null hypersurfaces. The outgoing code
would require some form of mesh refinement in order to resolve the
quasinormal ringdown for as many cycles as achieved by Papadopoulos.
However, the outgoing code avoids the late time caustic formation
noted in Papadopoulos's work, as well as the gauge ambiguity and
backscattering which complicate the extraction of waveforms on a
finite worldtube. An attractive option is to combine the best features
of both codes by matching an interior evolution based upon ingoing
null hypersurfaces to an exterior evolution based upon outgoing null
hypersurfaces, as implemented in~\cite{lehner} for spherically
symmetric Einstein-Klein-Gordon waves.

The waveform of relevance to gravitational wave astronomy is the
superposition of modes with different frequency compositions and
angular dependence.  Although this waveform results from a complicated
nonlinear processing of the input signal, which varies with choice of
observation angle, we have shown that the response of the individual
modes to an input signal of arbitrary amplitude can be obtained by
scaling the response to an input of standard reference amplitude. This
offers an economical approach to preparing a waveform catalog.

Work is in progress pursuing the above projects, as well as extending
the treatment to spinning black holes. In concert with other
approaches to compute nonlinear waveforms, it is hoped that robust
features of the gravitational waves produced by highly distorted black
holes will be discovered which could be exploited in data analysis
efforts. 

\begin{acknowledgments}

Discussions with L.~S. Finn spurred part of the present work. We thank
P. Brady and J. Pullin for comments and suggestions.  We thank the Center for
Gravitational Wave Physics at the Pennsylvania State University and
the Kavli Institute for Theoretical Physics at the  University of
California at Santa Barbara for their hospitality. This research was
supported by  National Science Foundation Grants PHY-9988663 to the
University of Pittsburgh, PHY-0135390 to Carnegie Mellon University
and PHY-9907949 to the  University of California at Santa Barbara. LL
was supported in part by the Horace Hearne Jr. Institute for
Theoretical Physics. Some computations were carried out on the
high-performing supercomputing facilities within the Louisiana State
University's Center for Applied Information Technology and Learning,
which is funded through Louisiana legislative appropriations.  The
codes were parallelized using the Cactus Computational
Toolkit~\cite{cactus}.

\end{acknowledgments}

\appendix

\section{Spin weighted spherical harmonics}
\label{ap:spin}

The complex stereographic coordinate $z = q+ip$ covers the sphere with
two patches which overlap in a region containing the equator. In the
north patch, the stereographic coordinate is related to standard
spherical coordinates $(\theta,\phi)$ by $z_N = \tan \frac{\theta}{2}
e^{i \phi}$; and in the south patch, by $z_S=\cot \frac{\theta}{2}
e^{-i \phi}$. In the overlap region, the coordinates are related by
$z_S=1/z_N$. 

In the north patch, the ordinary spherical harmonics are given
by~\cite{stewart.book}
\begin{eqnarray}
  Y_{\ell m} &=& \sum_{n=\max(0,m)}^{\min(\ell,\ell+m)}
        (-1)^n \frac{z_N^n \bar z_N^{n-m}}{n! (\ell+m-n)!
        (\ell-n)!(n-m)!} \nonumber \\
        &&\times \frac{\sqrt{1+2 \ell}}{P^\ell}\,
        \ell !\sqrt{\frac{(\ell-m)!(\ell+m)!}{4 \pi}}.
\end{eqnarray}
where $P=1+q^2 +p^2$.  We give the explicit formulae and conventions
for the spin-weighted spherical harmonics used in our calculations in
order to avoid confusion with various conventions found in the
literature. The spin-weighted spherical harmonics are defined by
\begin{eqnarray}
  {}_sY_{\ell\,m} = &\sqrt{\frac{(\ell-s)!}{(\ell+s)!}}
   \,\eth^s Y_{\ell\, m}, &
  s>0 \ , \\
  {}_sY_{\ell\,m} = &(-1)^s \sqrt{\frac{(\ell+s)!}{(\ell-s)!}}
  \,\bar\eth^{-s} Y_{\ell\, m}, & s<0
\end{eqnarray}
in terms of the spin-weight raising and lowering operators $\eth$ and
$\bar\eth$ which act on a spin-weight $s$ function $f$ according to
\begin{equation}
  \eth f = P^{1-s} \partial_{\bar z} (f P^s),  \quad
  \bar \eth f = P^{1+s} \partial_{z} (f P^{-s}).
\end{equation}
They are given in the north patch by
\begin{eqnarray}
  {}_sY_{N\, \ell m} &=& \sum_{n=\max(0,s+m)}^{\min(\ell+s,\ell+m)}
         (-1)^n
         \frac{\sqrt{1+2 \ell}}{(1+z_N \bar z_N)^\ell}
          \nonumber \\
         &\times&\frac{z_N^n \bar z_N^{n-m-s}}
         {n!(\ell+m-n)!(\ell+s-n)!(n-m-s)!} \nonumber \\
         &\times&  \,
         \sqrt{\frac{(\ell-m)!(\ell+m)!(\ell-s)!(\ell+s)!}{4 \pi}}.
\end{eqnarray}

The value $f_S(z_S)$ of any spin-weight $s$ function $f$ in the south
stereographic coordinates is given in terms of its value $f_N(z_N)$ in
north stereographic coordinates by~\cite{competh}
\begin{equation}
   f_S(z_S) = \left(-\frac{z_S}{\bar z_S}\right)^s  f_N(1/z_S) .
\end{equation}
Consequently, the spin-weighted spherical harmonics in the south
stereographic patch are given by
\begin{eqnarray}
   {}_sY_{S\,\ell m} &=& \sum_{n=\max(0,s+m)}^{\min(\ell+s,\ell+m)} 
   (-1)^{s+n}\frac{\sqrt{1+2 \ell}}{(1+z_S \bar z_S)^\ell}
         \nonumber \\
         &\times&\frac{z_S^{s+\ell-n} \bar z_S^{m+\ell-n}}
                     {n! (\ell+m-n)!(\ell+s-n)!(n-m-s)!} \nonumber \\
         &\times& \,
         \sqrt{\frac{(\ell-m)!(\ell+m)!(\ell-s)!(\ell+s)!}{4 \pi}}.
\end{eqnarray}

The spin-weighted spherical harmonics obey the orthogonality condition
\begin{equation}
  \oint {}_sY_{\ell\,m} \,\,{}_s\bar Y_{\ell'\,m'}\, d\Omega =
\delta_{\ell \ell '} \delta_{m m'}, \label{eq:cm}
\end{equation}
where the integration is over the entire solid angle $\Omega=4\pi$ of
the unit sphere. The integral~(\ref{eq:cm}) can be re-expressed as
\begin{eqnarray}
      \oint {}_sY_{\ell\,m}\,\, {}_s\bar Y_{\ell'\,m'} &d\Omega &=  \int_N 
         \frac{4(\,_sY_{N\,\ell m})
         (\,_s\bar Y_{N\,\ell' m'})}{(1+q_N^2+p_N^2)^2}
        dq_N \,dp_N \nonumber \\
        &+&
        \int_S \frac{4({}_sY_{S\,\ell m})({}_s\bar Y_{S\,\ell' m'})}
              {(1+q_S^2+p_S^2)^2} dq_S \,dp_S 
        \label{eq:cmsum}
\end{eqnarray}
where the subscripts $N$ and $S$ refer to the north and south patches
respectively and $\int_N$ and $\int_S $ denote  integration over the
corresponding hemispheres.

Rather than decompose $J$ and the news in terms of the
${}_sY_{\ell\,m}$ we decompose these functions in terms of the
${}_sR_{\ell\,m}$ which are defined by
\begin{eqnarray}
  {}_sR_{\ell\,m} &=& \frac{1}{\sqrt{2}} \left[ {}_sY_{\ell\,m} +
   (-1)^m {}_sY_{\ell\,-m} \right] \mbox{\ \ for $m>0$} \nonumber \\
  {}_sR_{\ell\,m} &=& \frac{i}{\sqrt{2}} \left[ (-1)^m 
   {}_sY_{\ell\,m} -{}_sY_{\ell\,-m}\right]
    \mbox{\ \ for $m<0$} \nonumber \\
  {}_sR_{\ell\,0} &=& {}_sY_{\ell\,0}. \label{eq:realharmonics}
\end{eqnarray}
These superpositions of the ${}_sY_{\ell\,m}$ obey the orthogonality
condition
\begin{equation}
  \oint {}_sR_{\ell\,m}\,\,{}_s\bar R_{\ell'\,m'}\, d \Omega = \delta_{\ell
\ell'} \delta_{m m'}.
\end{equation}
The advantage of using the ${}_sR_{\ell\,m}$ as basis functions is
that in the linear regime there is no coupling between different $m$
modes. The news calculation contains terms of the form  $\eth^2
(\bar\eth^2 J + \eth^2 \bar J)$ which introduces linear order mode
coupling in the ${}_sY_{\ell\,m}$ basis on account of 
$$
\eth^2 {}_2Y_{\ell\,m} + \bar\eth^2 {}_2\bar Y_{\ell\,m} 
 \sim \Re(Y_{\ell\,m})
\sim Y_{\ell\,m} + (-1)^m Y_{\ell\,-m}.
$$
However, since $\Re(R_{\ell\,m}) = R_{\ell\,m}$ there is no spurious
$m$ mode coupling in the ${}_sR_{\ell\,m}$ basis.

Any smooth spin-weight $s$ function $F$  can be decomposed into the
corresponding spin-weighted spherical harmonics according to
\begin{equation}
  F = \sum_{\ell=s, m=-\ell}^{\ell=\infty,m=\ell}
   F_{\ell\, m} \,\,{}_sR_{\ell\,m},
  \label{eq:decomp}
\end{equation}
where
\begin{equation}
  F_{\ell\, m} = \oint F \,\, {}_s\bar R_{\ell\,m}\, d\Omega.
\end{equation}
The decomposition in Eq.~(\ref{eq:decomp}) defines a spin-zero
potential $f$ for the spin-weighted function $F$ given by
\begin{eqnarray}
   f = \sum_{\ell m} \sqrt{\frac{(\ell -s)!}
     {(\ell +s)!}} F_{\ell\, m} \,R_{\ell\, m}\,\,\, \mbox {for
     $s>0$}, \\
   f = -1^s \sum_{\ell m} \sqrt{\frac{(\ell +s)!}
   {(\ell -s)!}} F_{\ell\, m} \,R_{\ell\, m}\,\,\, \mbox
{for $s<0$},
\end{eqnarray}
where $R_{\ell\, m} = {}_0R_{\ell\,m}$, $F = \eth^s f$ for $s>0$, and
$F = \bar \eth^{-s}f$ for $s<0$. Of particular relevance are the
coefficients of the various modes of the Bondi news function given by
\begin{eqnarray}
   N_{\ell m}&=&\oint N \,_2\bar R_{\ell\, m}\, d\Omega \nonumber \\
        &=&
        \int_N \frac{4}{(1+q_N^2+p_N^2)^2}N_N
        \,_2\bar R_{N\,\ell\, m}\, dq_N \,dp_N \nonumber \\
        &+&
        \int_S \frac{4}{(1+q_S^2+p_S^2)^2}N_S
        \,_2\bar R_{S\,\ell\, m}\, dq_S \,dp_S.
        \label{eq:coefficients}
\end{eqnarray}

\section{Modifications to the news module}
\label{sec:news}

In order to calculate the news function in an inertial Bondi gauge one
needs to track the phase angle which rotates the complex dyad defined
using the coordinates of the PITT null code into the complex unit
sphere dyad defined with respect to an inertial Bondi coordinate
system on $\scri^+$ (see Eq's~(32)-(36) of \cite{high}).  This phase,
which we denote by $e^{i \delta}$, obeys a hyperbolic partial
differential equation in the non-inertial coordinates of the code
(Eq.~(36) of
\cite{high})
\begin{eqnarray}
  (\partial_u + L^A \partial_A) \delta &=& \frac{1}{2}
\Im\Big(\frac{\bar{J_{,u}}J}{K+1} +
     \frac{J\left(U \bar\eth\bar J + \bar U \eth \bar J\right)}{2(K+1)}
\nonumber \\
     &+&J \bar \eth \bar U + K \bar \eth U  + 2 U \bar z\Big).
\label{eq:deltaOld}
\end{eqnarray}
The solution of Eq.~\ref{eq:deltaOld} requires consistent boundary
data at the edges of the stereographic patches. Because $\delta$ is
not a pure spin-weighted function (as is evident in the $U \bar z$
term in Eq.~(\ref{eq:deltaOld})), the necessary cross-patch
interpolation rules are complicated. However, we have found that the
computation of $\delta$ can be simplified by recasting
Eq.~\ref{eq:deltaOld} as an ordinary differential equation along the
characteristics of the equation, which are the null generators of
$\scri^+$. In the inertial Bondi coordinates, this ordinary
differential equation takes the form
\begin{eqnarray}
  \frac{d \delta}{du} &=& \frac{1}{2}
\Im\Big(\frac{\bar{J_{,u}}J}{K+1} +
    \frac{J\left(U \bar\eth\bar J + \bar U \eth \bar J\right)}{2(K+1)}
\nonumber \\
    &+&J \bar \eth \bar U + K \bar \eth U  + 2 U \bar z\Big)\Big|_{y^A},
\label{eq:deltaNew}
\end{eqnarray}
where $z$ and $u$ are the non-inertial coordinates. In this
formulation no boundary data are required. Eq.~(\ref{eq:deltaNew}) is
integrated to second order accuracy via
\begin{equation}
  \delta^{n+1}_i = \delta^n_i + \frac{1}{2}du
      \left(\mbox{RHS}^{n}_i + \mbox{RHS}^{n+1}_i\right),
\end{equation}
where $\mbox{RHS}$ is the right-hand-side of Eq.~(\ref{eq:deltaNew}),
the index $i$ labels the null generators of $\scri^+$, and the index
$n$ indicates the time level. (Although the above modification has
been implemented in the present code, it is not necessary when
considering axisymmetric spacetimes. In that case $\Im(U\bar z)=0$ and
$\delta$ behave as ordinary spin-weight 0 functions.) For more details
on the numerical implementation, see~\cite{yosefthesis}.

\bibliography{references}

\end{document}